\documentclass[aps,pra,twocolumn]{revtex4-1}

\usepackage{latexsym}		
\usepackage{amsmath}
\usepackage{graphicx}		
\usepackage{graphics}
\usepackage[usenames]{color}
\usepackage{color}

\begin{document}

\newcommand{\diff}[2]{\frac{\partial #1}{\partial #2}}
\newcommand{\secdiff}[2]{\frac{\partial^2 #1}{\partial #2^2}}
\newcommand{\comment}[1]{{\bf \textcolor{red}{ #1}}}
\newcommand{\BDEcomment}[1]{{\bf \textcolor{blue}{ #1}}}
\newcommand{\NPMcomment}[1]{{\bf \textcolor{ForestGreen}{ #1}}}

\title{Born-Oppenheimer study of two-component few-particle
  systems under one-dimensional confinement}

\author{N.~P.~Mehta}
\affiliation{Department of Physics and Astronomy, Trinity University, San Antonio, TX 78212-7200}
\email[]{nmehta@trinity.edu}

\date{\today}

\begin{abstract}
The energy spectrum, atom-dimer scattering length, and atom-trimer
scattering length for systems of three and four ultracold atoms with
$\delta$-function interactions in one dimension are presented as a 
function of the relative mass ratio of the interacting atoms.  The
Born-Oppenheimer approach is used to treat three-body (``HHL'')
systems of one light and two heavy atoms, as well as four-body (``HHHL'') systems
of one light and three heavy atoms.  Zero-range interactions of
arbitrary strength are assumed between different atoms, but the heavy
atoms are assumed to be noninteracting among themselves.  Fermionic and bosonic
heavy atoms with both positive and negative parity are considered.
\end{abstract}

\pacs{}

\maketitle

\section{Introduction}
\label{Section:Intro}

Cold-atom experiments now have the ability to
simultaneously control the atom-atom scattering length and the trapping
geometry. Quantum gases with essentially zero-range interactions in
one-dimensional (1D) trap geometries have been 
realized~\cite{Greiner2001PRL,Gorlitz2001PRL,Tolra2004PRL,Kinoshita2004Science,Kinoshita2006Nature,Leanhardt2002PRL}.
At the same time, the variety of atomic species that have been trapped
and cooled, including all of the alkali metals, continues to grow, ranging in mass from
Hydrogen~\cite{Fried1998PRL} to Radium~\cite{Parker2012PRC}.
Moreover, quantum degenerate mixtures of atoms have been the subject
of several experiments related to, for example, the creation of a gas of
degenerate polar molecules.~\cite{Ni2008Science}, the observation of
heteronuclear Efimov states~\cite{Barontini2009PRL}, and the
realization of mixtures of alkali atoms with alkaline-earth-like atoms~\cite{Hara2011PRL}.  

These recent experimental advances were preceded by a large body of
literature on the few and many-body physics of strongly interacting 1D
systems~\cite{LiebLiniger1963PhysRev,McGuire1964JMP,
  Yang1967PRL,YangYang1969JMP,Dodd1970JMP,Thacker1974PRD}.  More
recent theory work includes the calculation of the 3-boson hyperradial
potential curves~\cite{Gibson1987PRA, Amaya-Tapia1997FBS,
  Mehta2005PRA}, three-body recombination rates and threshold
laws~\cite{Mehta2007PRA}, and benchmark quality hyperspherical
calculations of three-boson binding energies and scattering
amplitudes~\cite{Chuluunbaatar2006JPB}.  The three-body
problem for unequal masses has been studied in
free-space~\cite{Kartavtsev2009JETP} and in an optical lattice\cite{Orso2010PRL}.

Of particular relevance to the present study is the mass-dependent
calculation of atom-dimer (2+1) scattering lengths and three-body
binding energies performed in~\cite{Kartavtsev2009JETP}.  The calculations of~\cite{Kartavtsev2009JETP} incorporate all of
the adiabatic hyperspherical potential curves necessary for numerical
convergence.  Here, instead of the (in principle) exact adiabatic hyperspherical representation~\cite{Macek1968JPB}, we use the
Born-Oppenheimer approach.  For the three-body calculations presented here, the
accuracy of the Born-Oppenheimer factorization is studied by direct
comparison to the results of~\cite{Kartavtsev2009JETP}, and that
comparison gives some quantitative insight to the accuracy of the
four-body calculations that follow.

It should also be noted that the HHHL system for spin-polarized
heavy fermions in three-dimensions (3D) has been studied by
Castin \emph{et al.}~\cite{Castin2010PRL}.  They found that for heavy
fermions with $J^\Pi=1^+$ symmetry, an infinite set
of four-body states appears in the mass range $13.384 < m_H/m_L <
13.607$.  Castin \emph{et al.} argue that these states have Efimov
character, however there seems to be some debate in the literature. 
Other authors~\cite{Guevara2012PRL} have argued these
are truly new states with properties distinct from Efimov states.
The authors of~\cite{Guevara2012PRL} consider particles interacting
with attractive $1/r^2$ interactions, basing their model on a
Born-Oppenheimer calculation of the potential energy surface governing
the heavy-particle dynamics. Better establishing the
accuracy of the Born-Oppenheimer approximation for short-range
potentials could potentially play a role in the interpretation of
these calculations.  

The Born-Oppenheimer approach has been successfully
applied to cold-atom systems in optical lattices to 
study novel crystalline phases in Fermi
mixtures~\cite{Petrov2007PRL}. The authors of~\cite{Petrov2007PRL} note
that the large mass ratios needed to observe these crystalline phases
can be achieved with small filling factors by tuning the
effective mass for the heavy particles.  We note that such a scheme could potentially be
used to observe the tetramer states predicted in this work.

In this paper, we consider 1D systems of three and four particles in which one particle
is ``light'' (of mass $m_L=\beta m_H$ with $0<\beta<1$) in comparison to
the remaining ``heavy'' (mass $m_H$) particles.
We restrict our attention to cases of noninteracting
heavy particles ($a_{HH} \rightarrow \infty$).  Here, $a_{HH}$ is the 1D
heavy-heavy scattering length.  We denote the 1D heavy-light
scattering length simply by $a$.  For cylindrical harmonic traps
in which only the lowest transverse mode is significantly populated,
the 1D scattering length may be expressed in terms of the 3D s-wave scattering length $a_{3D}$ and the
transverse oscillator length $a_\perp$ by the Olshanii formula~\cite{Olshanii1998PRL,Bergeman2003PRL}:
\begin{equation}
\label{Olshanii}
a = -\frac{a_\perp^2}{2a_{3D}}\left(1-C \frac{a_{3D}}{a_\perp} \right),
\end{equation}
\\
where $C \approx 1.4603$. Eq.~(\ref{Olshanii}) incorporates the
effect of virtual transitions to excited transverse modes.  When
$a_\perp=C a_{3D}$, Eq.~(\ref{Olshanii}) predicts a ``confinement
induced resonance'' (CIR), and the 1D scattering length vanishes.  
 
The degree to which the renormalization of the 1D atom-atom scattering
length by Eq.~(\ref{Olshanii}) accounts for the quasi-1D nature of the
confinement in few-body calculations is not a trivial
question~\cite{Mora2005PRA,Mora2005PRL,Levinsen2014Arxiv}.  Fully quasi-1D few-body calculations are complicated by the fact that
cylindrical confinement breaks spherical symmetry, and the total
angular momentum of the three or four-body system is not a good
quantum number.  In this paper, we proceed under the assumption that
meaningful few-body observables may be calculated with purely 1D
$\delta$-function interactions, renormalized according to Eq.~(\ref{Olshanii}).

This paper is organized as follows.
In Section~\ref{HHLSector}, we calculate the Born-Oppenheimer
potential curve describing the effective heavy-heavy interaction as
mediated by the light particle.  The HHL bound-state spectrum and the
H-HL scattering length is calculated as a function of the
heavy-light mass ratio.  The accuracy of the Born-Oppenheimer
approximation is studied by comparison to the high-accuracy
calculation of~\cite{Kartavtsev2009JETP}.

In Section~\ref{HHHLSector}, we calculate the two-dimensional
potential energy surface describing the heavy-particle dynamics in the
HHHL system. The adiabatic wavefunction describing the light particle is governed by a
one-dimensional Schr\"odinger equation with three $\delta$-functions.
We choose coordinates such that for a given permutation of heavy
particles, the ordering of the $\delta$-functions along the
light-particle coordinate is fixed.  The resulting energy surface is
then used in a calculation of the three-body adiabatic hyperradial
potential curves for the heavy particles.  From those potential
curves, the HHHL binding energies and H-HHL scattering lengths are calculated.

\section{Three-Body (HHL) Problem}
\label{HHLSector}
Let particles 1 and 2 have mass $m_1 = m_2 = m_H$ and particle 3
have mass $m_L=\beta m_H$. Throughout this paper, we
set $\hbar = 1$.  For a zero-range heavy-light interaction of the form
$V_{ij} = g \delta(x_i - x_j)$, the 1D H-L scattering length is $a =
-1/(\mu_{\text{HL}}g)$, and assuming $a > 0$, the heavy-light binding energy is $B_2
= \mu_{\text{HL}} g^2/2 = 1/(2\mu_{\text{HL}} a^2)$. For particle positions
$\{x_1, x_2, x_3\}$, we introduce the following unitless mass-scaled
Jacobi coordinates (See Fig.~\ref{fig:JacobiTrees}):
\begin{align}
\label{HHLcoors}
 x &= \frac{1}{a} \sqrt{\frac{\mu_{12}}{\mu_{3b}}} (x_2 - x_1)\\
y &= \frac{1}{a} \sqrt{\frac{\mu_{12,3}}{\mu_{3b}}}\left(\frac{m_1 x_1 + m_2 x_2}{m_1+m_2}-x_3\right)
\end{align}
Here, $\mu_{12}=m_H/2$, $\mu_{12,3}=m_H [2\beta/(2+\beta)]$ and $\mu_{3b} =
\sqrt{\mu_{12} \mu_{12,3}}$ are reduced
masses.  
The heavy-light reduced mass is $\mu_{\text{HL}} = m_H [\beta/(1+\beta)]$.
It is convenient to scale the Hamiltonian by the heavy-light binding energy:
\begin{equation}
\label{B2}
B_2 = \frac{1}{m_H a^2}\frac{\beta +1}{2 \beta},
\end{equation}
so that all energies are measured in units of $B_2$.  The
Schr\"odinger equation then reads:
\begin{widetext}
\begin{equation}
\label{HHLSE}
-\frac{1}{2\mu_3}\left(\secdiff{}{x}+\secdiff{}{y}\right)\Psi(x,y)+g_3\left(\lambda
  \delta (2 x_0) + \delta(y+x_0) + \delta(y-x_0)
\right)\Psi(x,y)=E \Psi(x,y).
\end{equation}
\end{widetext}
The parameter $\lambda$ is the ratio of the heavy-heavy coupling to
the heavy-light coupling.  In this work, only $\lambda \rightarrow 0$
and $\lambda \rightarrow \infty$ are considered.  The notational cost
of scaling by $B_2$ is contained in the definition of the following
unitless parameters:
\begin{align}
\mu_3 &= \frac{1+\beta}{2\sqrt{\beta(2+\beta)}} \\
g_3 &= -2\sqrt{2} \left(\frac{\beta}{2+\beta}\right)^{1/4} \\
x_0 &= x \sqrt{\frac{\beta}{2+\beta}}  
\end{align}
We now assume the wavefunction may be approximated by the Born-Oppenheimer product:
\begin{equation}
\label{BOanzats}
\Psi(x,y) = \Phi(x;y)\psi(x)
\end{equation}
where $\Phi(x;y)$ is a solution to the fixed-$x$ equation,
\begin{equation}
\label{adiabaticHHLSE}
\begin{split}
&\left[\frac{-1}{2\mu_3}\secdiff{}{y} +
g_3\left(\delta(y+x_0) + \delta(y-x_0) \right)
\right] \Phi(x;y) \\
&= u(x) \Phi(x;y),
\end{split}
\end{equation}
and $u(x)<0$ is the Born-Oppenheimer potential in units of the H-L binding energy. 
Note that the solutions $\Phi(x;y)$ and the potential curve $u(x)$ are
independent of $\lambda$.  Inserting Eq.~(\ref{BOanzats}) into
Eq.~(\ref{HHLSE}) and making use of Eq.~(\ref{adiabaticHHLSE}) yields,
\begin{equation}
\label{HHEffEq}
\left(\frac{-1}{2\mu_3}\secdiff{}{x}+g_3\lambda\delta(2x_0)+u(x)+\frac{\tilde{Q}(x)}{2\mu_3}\right)\psi(x)=E\psi(x)
\end{equation}
where,
\begin{equation}
\label{Q11}
\tilde{Q}(x) = \left \langle \diff{\Phi}{x}\right. \left |
    \diff{\Phi}{x} \right \rangle_y 
\end{equation}
It is understood that the integration in the matrix element $\tilde{Q}(x)$ is
carried out over the $y$ coordinate only, while the adiabatic
coordinate $x$ is held fixed.
\subsection{Solution to The Adiabatic Equation}
\label{AdiabaticHHL}

Equation~(\ref{adiabaticHHLSE}) is symmetric with respect to the
operation $y \rightarrow -y$, and so the eigenstates $\Phi(x;y)$ must
be even or odd under that operation.  The elementary solutions that
vanish as $|y|\rightarrow\infty$ are conveniently written for
positive $y$ as:
\begin{equation}
\label{HHLsolns}
\Phi(x;y) = 
\begin{cases}
A \sinh(\kappa y) + B \cosh(\kappa y),\; & \text{if $0 \le y \le x_0$;}\\
D e^{-\kappa y},\; & \text{if $x_0 \le y$;}
\end{cases}
\end{equation}
where $\kappa(x)=\sqrt{-2\mu_3u(x)}$. For the even solution, $A=0$, while
for the odd solution, $B=0$.  Matching the wavefunctions, and imposing
the derivative discontinuity across the delta-function at $y=x_0$
leads to the following transcendental equation for the eigenvalue $\kappa$:
\begin{equation}
\label{HHLTEQ}
\frac{\kappa }{g_3 \mu_3}+1=(-1)^{P+1} e^{-2 \kappa  x_0}.
\end{equation}
Here, $P=0$ corresponds to the (even) solution for which
$\left.\diff{\Phi}{y}\right|_{y=0}=0$, and $P=1$ corresponds to the
(odd) solution for which $\Phi|_{y=0}=0$. Borrowing language from
molecular physics, one can view the $P=0$ solution as
belonging to the ``bonding'' orbital, and the $P=1$ solution to the
``anti-bonding'' orbital.  


The potential curves resulting from the $x$-dependent solution to
Eq.~(\ref{HHLTEQ}) for $\beta^{-1}=22.08$ (for Li-Cs mixtures) are shown in
Fig.~\ref{fig:LiCsCurves}(b)~and~\ref{fig:LiCsCurves}(d).  The potential curves shown in these
two graphs are identical because Eq.~(\ref{HHLSE}) is independent of
the heavy-particle symmetry.  Any apparent differences are due to the energy scales on
the graph.  The bound-state structure, however, \emph{is} dependent on
the heavy symmetry through the boundary condition placed on $\psi(x)$
at $x=0$.  Note that the $\lambda = \infty$ solutions to
Eq.~(\ref{HHEffEq}) for heavy bosons are identical to those for
\emph{noninteracting} heavy fermions. The boundary condition
$\psi(0)=0$ is applied for fermionic
heavy atoms as well as \emph{fermionized} bosonic
atoms, leading to the correspondence first recognized in~\cite{Girardeau1960JMP}.
For small heavy-atom separations, there is no $P=1$ negative
energy solution to Eq.~(\ref{HHLTEQ}), and the light particle
is lost to the continuum where the excited state potential
terminates at the zero-energy threshold.

\begin{widetext}
\setlength{\unitlength}{1in}
\begin{center}
\begin{figure}[!t]
\leavevmode
\includegraphics[width=6.2in,clip=true]{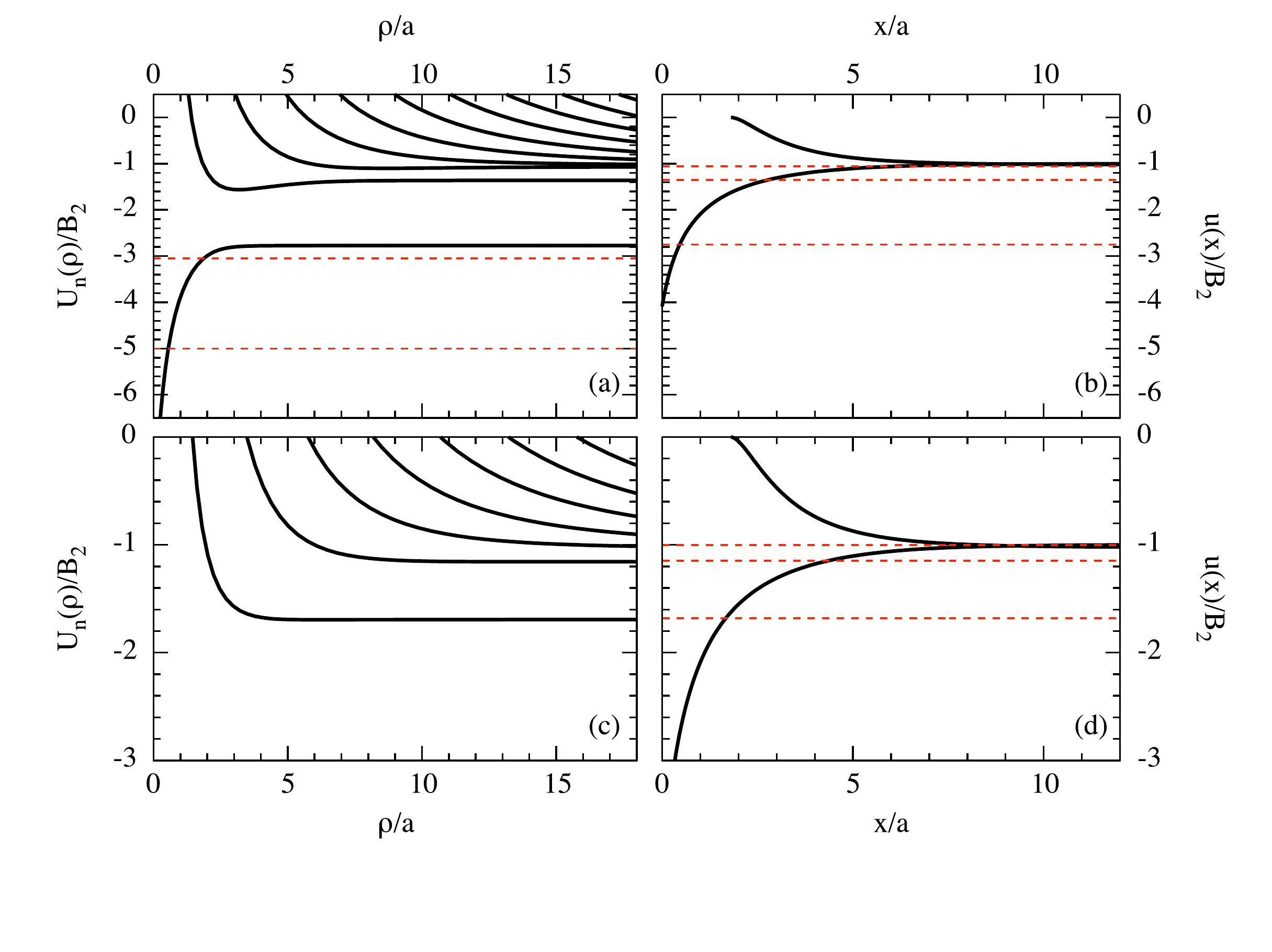}
\caption{(color online) Graph (a) shows the bosonic hyperradial
  potential curves describing the heavy-particle dynamics in the HHHL system, while graph
  (b) shows the corresponding Born-Oppenheimer potential curves for
  the HHL system.  Graphs (c) and (d) show similar curves, except for
  fermions.  All graphs are for $\beta^{-1} = 22.08$,
  appropriate for Li-Cs mixtures.  The dashed-red lines indicate bound
  states.}
\label{fig:LiCsCurves}
\end{figure}
\end{center}
\end{widetext}

The atom-dimer scattering length and the HHL spectrum are to a very
good approximation determined solely by the potential curve corresponding to
the $P=0$ solution to Eq.~(\ref{HHLTEQ}), so we shall restrict
our immediate focus to that solution.
When the heavy atoms are far apart, $|y-x_0| \gg |x_0|$,
Eq.~(\ref{HHLSE}) behaves as though there is a single $\delta$-function of modified strength
at the origin.  The solutions to
Eq.~(\ref{HHLTEQ}) when $x_0\rightarrow \infty$ underestimate the
correct threshold energy by $\beta/(2+\beta)$:
\begin{equation}
\label{ulargex}
\lim_{|x|\to \infty}{u(x)}= -1-\frac{\beta}{2+\beta}
\end{equation}
This result is not unexpected, since we have so far neglected the
positive-definite contribution $\tilde{Q}(x)/(2\mu_3)$ to the heavy-particle
kinetic energy.  It is known that neglecting this ``diagonal
correction'' --- which we call the ``Extreme Adiabatic Approximation'' (EAA) --- yields a lower bound $E_{EAA}$ to the $N$-body bound-state
energy.  Including the diagonal correction, but neglecting any
couplings between Born-Oppenheimer curves --- an approximation we
call the ``uncoupled adiabatic approximation'' (UAA) --- yields an
upper bound $E_{UAA}$ to
the correct energy~\cite{Brattsev1965Dokl,Epstein1966JCP,Strace1979PRA,Coelho1991PRA}.
We find for this problem that the trend in these
inequalities $E_{EAA}<E<E_{UAA}$ is already present in the threshold values of the
adiabatic potential itself.  In other words, we find that in the limit
$|x|\rightarrow \infty$, $u(x)<-1<u(x)+\tilde{Q}(x)/(2\mu_3)$.  In the next
section, we explicitly calculate $\tilde{Q}(x)$.

\subsection{The Diagonal Correction $\tilde{Q}/(2\mu_3)$}
\label{}
Using the solutions Eq.~(\ref{HHLsolns}) (with $A=0$) along with the
normalization Eq.~(\ref{HHLnorm}), we explicitly calculate the integral
involved in the nonadiabatic correction Eq.~(\ref{Q11}).  Taking $A=0$
in Eq.~(\ref{HHLsolns}), continuity of the wavefunction immediately
yields $D = B\cosh(\kappa  x_0)/e^{-\kappa x_0}$.  The remaining
normalization constant, $B$, depends on the H-H separation distance
both explicitly, and implicitly through the eigenvalue $\kappa$:
\begin{equation}
\label{HHLnorm}
B(x)=\frac{2 \sqrt{\kappa}}{\sqrt{2 x_0 \kappa + e^{2 x_0 \kappa} + 1}}
\end{equation} 
Derivatives of the eigenvalue $\kappa(x)$ are replaced by
expressions involving $\kappa$ itself by differentiating
Eq.~(\ref{HHLTEQ}) with respect to $x$ and solving for $\kappa'$.  
We find that the nonadiabatic correction $\tilde{Q}(x)$ can be expressed as a rational
polynomial in the separation distance $x$: 
\begin{align}
\label{Q11full}
&\tilde{Q}(x)=\frac{\beta}{3(2+\beta) \left(-2 h x_0+2 \kappa  x_0+1\right){}^4}\notag\\
&\times\big[3 h^2+x_0 \left(-12 h^3+24 h^2 \kappa -36 h \kappa ^2+24 \kappa ^3\right) +\notag\\
& x_0^3 \left(-16 h^3 \kappa ^2+48 h^2 \kappa ^3-48 h \kappa ^4+16 \kappa ^5\right)+\notag\\
&x_0^2 \left(12 h^4-48 h^3 \kappa +108 h^2 \kappa ^2-120 h \kappa ^3+48 \kappa ^4\right)+\notag\\
&x_0^4 \left(-16 h^4 \kappa ^2+64 h^3 \kappa ^3-96 h^2 \kappa ^4+64 h \kappa ^5-16 \kappa ^6\right)\big],
\end{align}
where we have defined the constant $h=\mu_3g_3$.  Evaluating
Eq.~(\ref{Q11full}) at the asymptotic value of the potential
Eq.~(\ref{ulargex}) gives $\tilde{Q}/2\mu_3 \rightarrow [\beta/(2+\beta)]+[\beta/(2+\beta)]^2$, and including $\tilde{Q}/(2\mu_3)$ in Eq.~(\ref{HHEffEq}) yields the correct
threshold energy to order $(\frac{\beta}{2+\beta})^2$:
\begin{equation}
\label{adthreshold}
\lim_{|x|\rightarrow \infty}{\left[ u(x)+\frac{\tilde{Q}}{2\mu_3}\right]} =
-1 + \left(\frac{\beta}{2+\beta}\right)^2.
\end{equation}
In other words, for small $\beta$, the error in the threshold energy vanishes linearly without the
diagonal correction, but quadratically when it is included.
Interestingly, for the equal mass case ($\beta = 1$), the UAA gives the correct threshold
energy to within 11\%.  This may seem a somewhat
surprising result since the Born-Oppenheimer factorization is
typically expected to fail catastrophically in this limit, however other
authors~\cite{Adhikari1992NuovoCimento} have found the
Born-Oppenheimer approach to work surprisingly well for
short-range s-wave interactions in 3D for a wide variety of mass
ratios.  It seems that the present 1D calculation shares similar good-fortune.



\subsection{Numerical results for the HHL system}
Here, we compare the present Born-Oppenheimer calculation for the HHL
system to the high-accuracy calculations
of~\cite{Kartavtsev2009JETP}.  Binding energies and scattering
solutions are calculated in the UAA.  

For the scattering calculation, $u(x)$ and
$\tilde{Q}(x)$ are calculated to 15 digits on a uniform grid, and
the Numerov method is used to propagate the solution out from $x=0$ to
some $x_{max}$.  The attractive well in $u(x)$ widens as the mass
ratio $\beta^{-1}$ increases.  An $x_{max} \sim 40$ is
sufficient for $\beta^{-1} \lesssim 10$, but must be increased to $x_{max}
\sim 120$ for $\beta^{-1} \sim 250$.  For a Numerov step size $s$,
each integration step in the Numerov method
can introduce an error of order $s^5$.  For $N_s$ total steps, an upper
bound to the asymptotic values of the wavefunction of order $N_s s^5 =
x_{max} s^4$ is maintained less than $10^{-10}$.  The asymptotic wavefunction is matched to:
\begin{equation}
\label{matching}
\psi(x)\rightarrow 
\begin{cases}
\cos(kx) - \tan(\delta)\sin(kx) \;\;\; \text{Bosons} \\
\sin(kx) + \tan(\delta)\cos(kx) \;\;\; \text{Fermions} 
\end{cases}
\end{equation}
The atom-dimer scattering length is then extracted from the effective
range expansion as:
\begin{equation}
\label{ere}
\frac{1}{a_{AD}} = \lim_{k_{AD}\rightarrow 0}
\begin{cases}
k_{AD}\tan{\delta} \;\;\; \text{Bosons} \\
-k_{AD}\cot{\delta} \;\;\; \text{Fermions}
\end{cases}
\end{equation}
Here, $k_{AD} = \sqrt{2 \mu_{23,1} E_{rel}}$, while $k=\sqrt{2\mu_3 E_{rel}}$.
The mass ratios in Table~\ref{table:ScatLengths} (discussed below) are obtained by a
bisection root-finding algorithm (either on $1/a_{AD}$ or on $a_{AD}$)
to 6-digit precision.  The number of digits reported here
represents the precision of our calculation.  The accuracy is
best estimated by comparing to the calculations of~\cite{Kartavtsev2009JETP}.

Bound-state calculations are performed variationally by expanding
$\psi(x)$ in a basis of b-splines, and
solving the resulting generalized eigenvalue problem.  We have
verified that the results are well converged with respect to the
number of grid points used to interpolate the potential $u(x)$, as well as the
number and placement of b-splines. 

\begin{figure}[h]
\begin{center}
\leavevmode
\includegraphics[width=3.5in,clip=true]{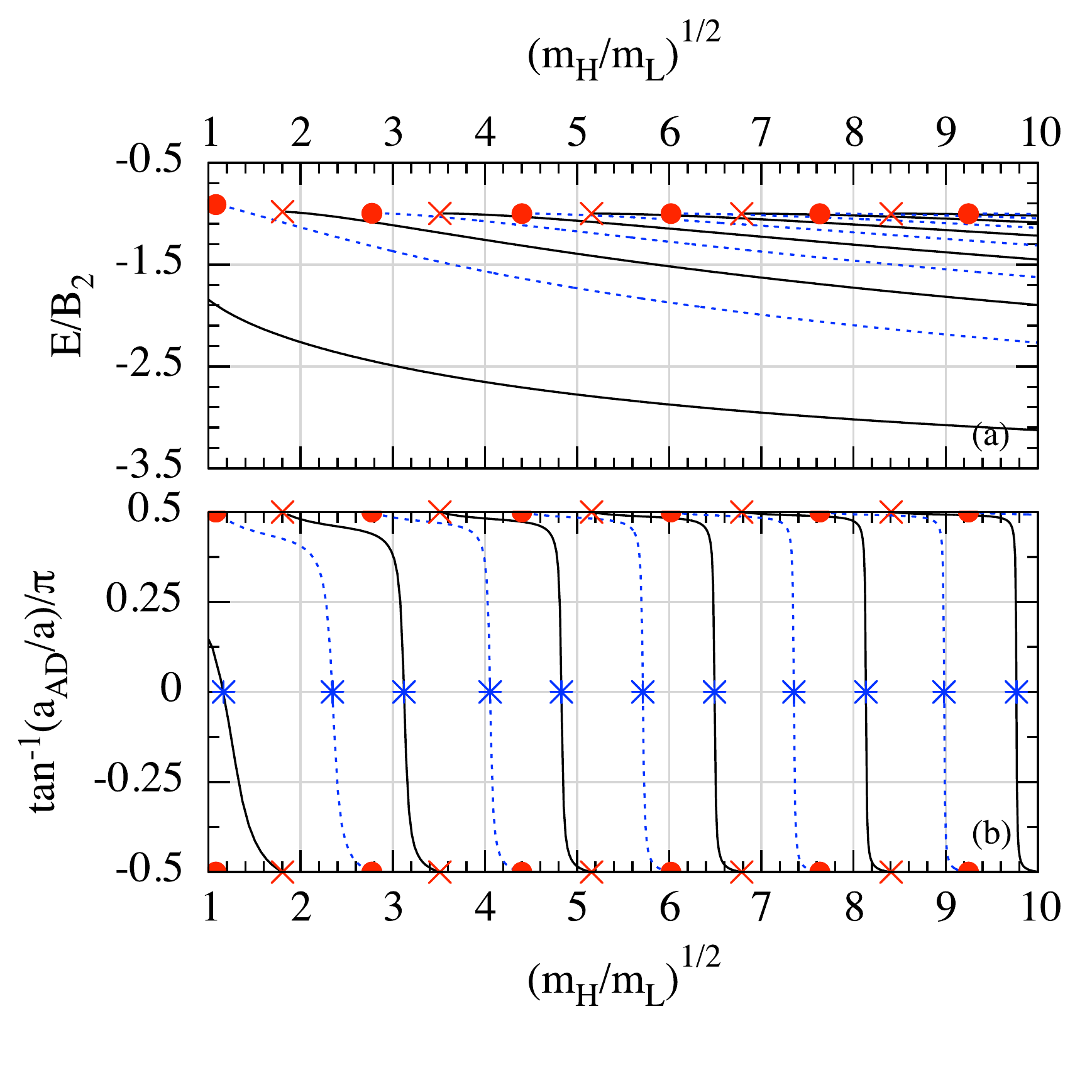}
\caption{Graph (a) shows the energy spectrum for the three-body (HHL) system as a function $\beta^{-1/2}$. Graph (b) shows
  $\tan^{-1}{(a_{AD}/a)}/\pi$, where $a_{AD}$ is the atom-dimer
  scattering length.  For both graphs, the
  solid-black curves denote noninteracting bosonic H-particles.  The
  dashed-blue curves denote noninteracting fermionic particles, or 
  equivalently, \emph{fermionized} H-particles with
  $\lambda \rightarrow \infty$.  Red crosses and dots, for bosons and
  fermions respectively, indicate the appearance
  of a new bound state and $|a_{AD}|\rightarrow \infty$.  Blue stars
  indicate $a_{AD}\rightarrow 0$.}
\label{fig:HHLSpectrum}
\end{center}
\end{figure}

In Fig.~\ref{fig:HHLSpectrum}(a), we show the three-body spectrum as a
function of $\beta^{-1/2}$ (recall $\beta= m_L/m_H$).  The mass ratios at
which a new state appears, marked by the red crosses
for $\lambda = 0$ and red dots for $\lambda = \infty$, trace out a
curve governed by the $\beta$-dependence of the threshold Eq.~(\ref{adthreshold}).  In the hyperspherical calculation of~\cite{Kartavtsev2009JETP}, the
threshold is reproduced exactly, and all dots and crosses appear at
$E_2/B_2 = -1$.
\setlength{\unitlength}{1in}

\begin{table}[!t]
  \caption{
    \label{table:ScatLengths}
    The values of the mass ratio $\beta^{-1}=m_H/m_L$ for which the
    atom-dimer scattering length is infinite
    ($a_{AD}\rightarrow\infty$, corresponding to the appearance of
    the $n^{\text{th}}$ trimer state), or zero ($a_{AD}\rightarrow
    0$), are tabulated both the
    case of noninteracting bosonic H atoms ($\lambda \rightarrow 0$)
    and fermionic H atoms ($\lambda \rightarrow \infty$).
    Results are compared to Ref.~\cite{Kartavtsev2009JETP}.  An asterisk (*) denotes an exact
    result.}
  \begin{center}
    \begin{tabular}{|c|c|c|c|c|}
      \hline
         &\multicolumn{2}{c|}{$\lambda = 0$}&\multicolumn{2}{c|}{$\lambda = 0$}\\
      $n$&\multicolumn{2}{c|}{$\beta^{-1}\large|_{a_{AD}\rightarrow 0}$} &\multicolumn{2}{c|}{$\beta^{-1}\large|_{a_{AD}\rightarrow  \infty}$}\\
      \cline{2-5}
      &This work&Ref.~\cite{Kartavtsev2009JETP}&This work&Ref.~\cite{Kartavtsev2009JETP} \\
      \hline
      1  &  $-$    &   $-$        &$-$      &       $-$         \\
      2  & $1.357$    & $0.971$   &$3.255$&$2.86954$        \\
      3  &$9.747$& $9.365$    &$12.336$&$11.9510$      \\
      4  &$23.333$& $22.951$  &$26.602$&$26.218$        \\
      5  &$42.142$& $41.762$  &$46.055$&$45.673$      \\
      6  &$66.168$& $65.791$  &$70.695$&$70.317$       \\
      7  &$95.404$& $95.032$  &$100.523$&$100.151$    \\
      8  &$129.845$&$129.477$ &$135.539$&$135.170$    \\
      9  &$169.488$&$169.120$ &$175.742$&$175.374$    \\
      10 &$214.331$& $213.964$&$221.133$&$220.765$     \\
      \hline
    \end{tabular}
    \begin{tabular}{|c|c|c|c|c|}
      \hline
      &\multicolumn{2}{c|}{$\lambda = \infty$}&\multicolumn{2}{c|}{$\lambda = \infty$}\\
      $n$&\multicolumn{2}{c|}{$\beta^{-1}\large|_{a_{AD}\rightarrow 0}$} &\multicolumn{2}{c|}{$\beta^{-1}\large|_{a_{AD}\rightarrow  \infty}$}\\
      \cline{2-5}
      &This work&Ref.~\cite{Kartavtsev2009JETP}&This work&Ref.~\cite{Kartavtsev2009JETP} \\
      \hline
      1 &$-$      &     $0^*$   & $1.170$  &     $1^*$                         \\            
      2 &$5.499$ &$5.2107$     &$7.694$& $7.3791$ \\            
      3 &$16.456$&$16.1197$  &$19.373$&$19.0289$\\         
      4 &$32.650$&$32.298$   &$36.235$&$35.879$ \\          
      5 &$54.067$&$53.709$   &$58.283$&$57.923$ \\           
      6 &$80.697$&$80.339$   &$85.518$&$85.159$ \\          
      7 &$112.535$&$112.179$ &$117.940$&$117.583$ \\        
      8 &$149.577$&$149.222$ &$155.550$&$155.193$ \\      
      9 &$191.820$&$191.463$ &$198.347$&$197.989$ \\       
      10&$239.262$&$238.904$ &$246.331$&$245.973$ \\
      \hline                      
    \end{tabular}
  \end{center}
\end{table}

Figure~\ref{fig:HHLSpectrum}(b) shows $\tan^{-1}{(a_{AD}/a)}/\pi$ as a
function of $\beta^{-1/2}$.  Again, the red dots and crosses denote
the mass ratios at which a new state appears and the atom-dimer
scattering length $a_{AD}\rightarrow\infty$.  The blue stars indicate
$a_{AD}\rightarrow 0$.  In a manner similar
to~\cite{Kartavtsev2009JETP}, we tabulate these particular values of
the mass ratio in Table~\ref{table:ScatLengths}.  Note that the Born-Oppenheimer
calculation consistently overestimates the critical mass ratios $\beta^{-1}$ by
approximately $0.3-0.4$.  The overestimate is understood, at least
qualitatively, by noting that the $UAA$ produces an upper bound to the
binding energy, and the trend in the spectrum is for deeper binding as
$\beta^{-1}$ increases.  The percentage error in the critical values
of $\beta^{-1}$ decreases monotonically, as one might expect.

The $\beta = 1$ HHL ground state for $\lambda = 0$ bosons was found
in~\cite{Kartavtsev2009JETP} to be (in units of $B_2$) $E_3 = -2.087719$, very
close to the value $E_3 = -2.08754$ found much earlier
in~\cite{GaudinJdePhys1975}.  Here, we find that the EAA produces a lower
bound of $E_{3,EAA} = -2.4227$, approximately 16\% deeper than
the correct value.  The UAA underbinds by about 11\%, giving
the upper bound $E_{3,UAA}=-1.8561$.  It is interesting that the
error in the UAA calculation is almost entirely accounted for by the
overestimate of the atom-dimer threshold energy.  In fact, scaling by the
threshold energy Eq.~(\ref{adthreshold}), one obtains $E_{3,UAA}/E_{\text{thresh}} = -2.0879$,
overbinding by only $0.01\%$.

Kartavtsev and Malykh~\cite{Kartavtsev2007JPhysB} found that universal (non-Efimov)
fermionic states in 3D exist for mass ratios $\beta^{-1} \gtrsim
8.17$.  Pricoupenko and Pedri~\cite{Pricoupenko2010PRA} found similar
states in 2D for $\beta^{-1}\gtrsim 3.33$.  Levinsen and
Parish~\cite{Levinsen2013PRL} established that these states are
continuously connected as confinement is increased.  It is interesting to
speculate whether the fermionic state that appears in 1D at
$\beta^{-1} = 1$ ($\beta^{-1}\approx 1.170$ in our calculation) is
continuously connected to these universal trimer states in higher 
dimensions.  

\section{Four-Body (HHHL) Problem}
\label{HHHLSector}

\begin{figure}[t]
\begin{center}
\leavevmode
\includegraphics[width=2.5in,clip=true]{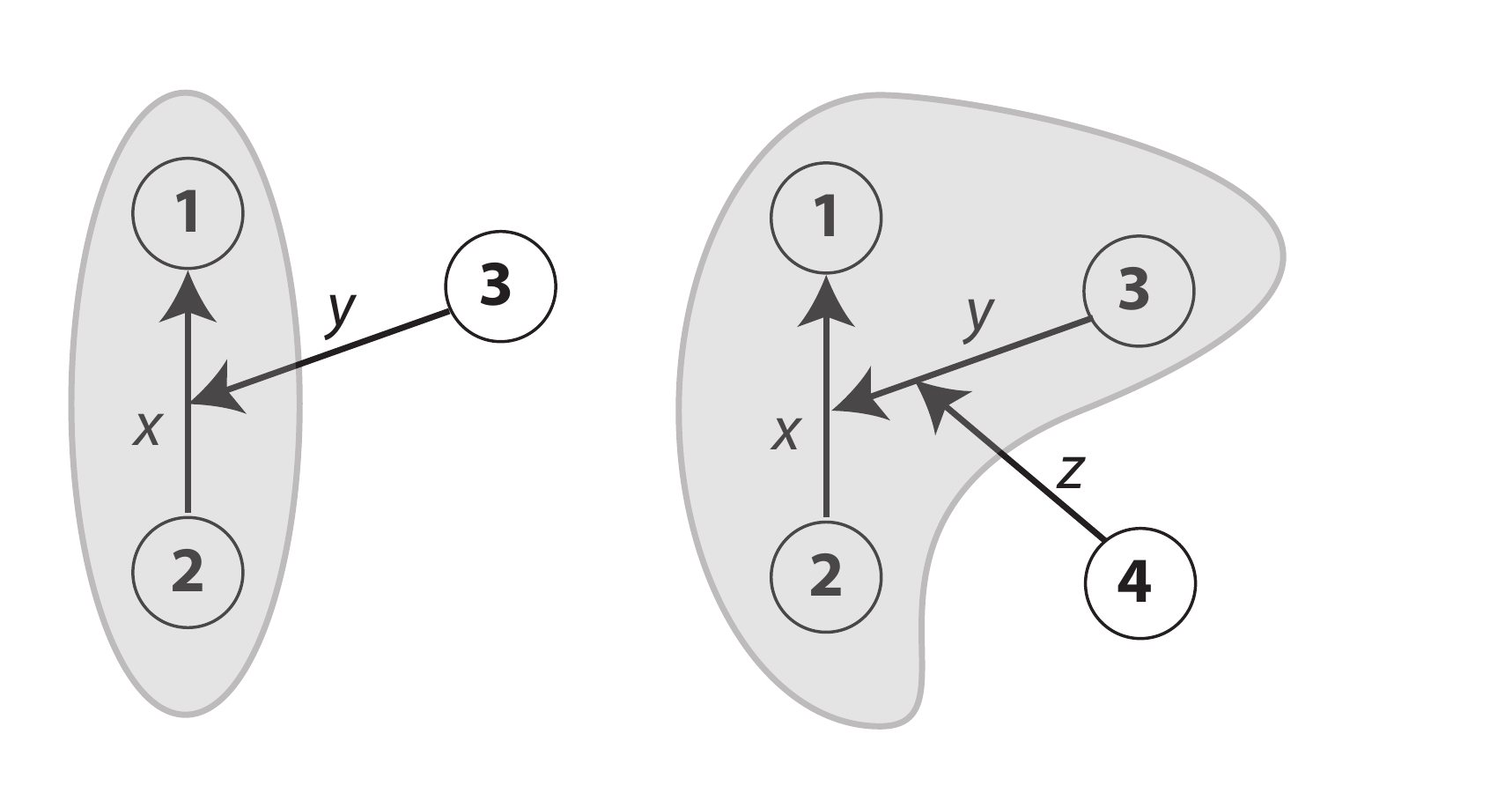}
\caption{A schematic diagram of the Jacobi coordinates for (a) the
  three-body problem and (b) the four-body problem are shown.
  The heavy particles are contained in the shaded regions.}
\label{fig:JacobiTrees}
\end{center}
 \begin{picture}(0,0)(0,0)
      \put(-0.6,1.3){(a)}
      \put(0.8,1.3){(b)}
 \end{picture}
\end{figure}
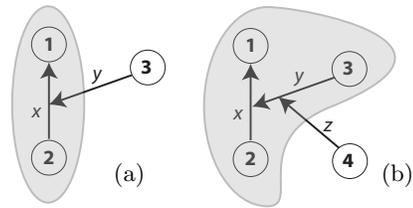

Let us now turn to the calculation of four-body observables.  The
basic three-step recipe for this calculation is as follows. First, the
Born-Oppenheimer method is used to calculate the 2D potential energy
surface for the heavy particles in the extreme adiabatic
approximation (EAA).  Next, this potential energy surface is inserted into a
calculation of the hyperradial adiabatic potential curves and
couplings.  Finally, the resulting set of coupled hyperradial
equations is solved for the bound-states and atom-trimer scattering
length.  The entire procedure is then repeated for different values of
$\beta$. If a sufficiently large number of hyperradial curves and couplings are
included in the final step, then the accuracy of the calculation is
limited almost entirely by the EAA made in the first step.

\subsection{The adiabatic equations}
For all four-body (HHHL) calculations that follow, we choose particles
1, 2 and 3 to have mass $m_1=m_2=m_3=m_H$ and particle 4 to have mass $m_4=\beta
m_H$.  The solution to the adiabatic equation is most easily carried
out using the ``K-type'' Jacobi coordinates shown in
Fig.~\ref{fig:JacobiTrees}(b), with unitless mass-scaled coordinates defined as:
\begin{align}
\label{JacobiCoors}
x &= \frac{1}{a}\sqrt{\frac{\mu_{12}}{\mu_{4b}}}\left(x_1 - x_2 \right) \notag \\
y &= \frac{1}{a}\sqrt{\frac{\mu_{12,3}}{\mu_{4b}}}\left(\frac{m_1 x_1 + m_2 x_2}{m_1+m_2}-x_3 \right) \notag\\
z &= \frac{1}{a}\sqrt{\frac{\mu_{123,4}}{\mu_{4b}}}\left(\frac{m_1 x_1 + m_2 x_2 + m_3 x_3}{m_1 + m_2 + m_3}-x_4 \right)
\end{align}
Here, $\mu_{4b}=(\mu_{12}\mu_{12,3}\mu_{123,4})^{1/3}$ is the
four-body reduced mass.  Again, we rescale the Schr\"odinger equation
by the heavy-light binding energy $B_2$.  The full four-body
Schr\"odinger equation then reads:
\begin{equation}
\begin{split}
\frac{-1}{2 \mu_4}&\nabla ^2\Psi(\rho,\phi,z) + \Big[  g_4\sum_{i=1}^{3}{\delta(z-z_i)} \\  
+ &\lambda g_4 \sum_{i<j}^{3}{\delta\left(\alpha
      \rho\left|\sin(\phi-\phi_{ij})\right|\right)} \Big]
\Psi(\rho,\phi,z) =E \Psi(\rho,\phi,z) 
\end{split}
\end{equation}
where $\alpha = \sqrt{6}[(3+\beta)/\beta]^{1/6}$, $\phi_{12}=\pi/2$,
and $\phi_{23}=-\phi_{13}=\pi/6$. Again, rescaling by $B_2$ introduces
the following unitless parameters:
\begin{align}
\mu_4 &= \frac{\beta +1}{2 \beta ^{2/3} (3+\beta)^{1/3}}\\
g_4 &= -2 \sqrt{3} \left({\frac{\beta }{3+\beta}}\right)^{1/3} \\
z_i&=\sqrt{\frac{2\beta}{3+\beta}} \; \rho \sin \left(\phi - \phi_i\right) 
\end{align}
where $\phi_1 = -4\pi/3$, $\phi_2 = 0$, and $\phi_3 = -2\pi/3$.
The particular choice of Jacobi coordinates
Eq.~(\ref{JacobiCoors}) has the advantage that the separation distances
$x_{12}$, $x_{13}$ and $x_{23}$ are all independent of the
$z$-coordinate.  The heavy-particle dynamics is restricted to the
$x$-$y$ plane, and the light particle can be integrated out by solving
an equation in the $z$-coordinate only, with \emph{fixed} $x$ and $y$.
The transformation to hyperspherical coordinates is accomplished by
expressing $x$, $y$, and $z$ in terms of the usual spherical polar
coordinates $R$, $\theta$ and $\phi$.  The
heavy-particle subsector is then described by $x=\rho\cos\phi$ and
$y=\rho\sin\phi$, where $\rho=\sqrt{x^2+y^2}$ is the projection of $R$ onto the
$x$-$y$ plane: $\rho=R\sin\theta$, and $z=R\cos\theta$.  

Clearly, fixing $x$ and $y$ is equivalent to fixing $\rho$ and
$\phi$.  We make the Born-Oppenheimer factorization:
\begin{equation}
\label{BOfact}
\Psi = \Phi(\rho,\phi;z)\psi(\rho,\phi),
\end{equation}
where the adiabatic equation for the Born-Oppenheimer surface is:
\begin{equation}
\label{HHHLAdSE}
\begin{split}
\left(\frac{-1}{2\mu_4}\secdiff{}{z}+g_4\sum_{i=1}^{3}{\delta(z-z_i)}\right)\Phi(\rho,\phi;z)=U(\rho,\phi)\Phi(\rho,\phi;z)
\end{split}
\end{equation}
The heavy-particle eigenstates now live on the potential energy
surface $U(\rho,\phi)$, and satisfy (in the EAA):
\begin{equation}
\label{BOEquation}
\begin{split}
&\frac{-1}{2\mu_4}\left(\frac{1}{\rho}\diff{}{\rho}\rho\diff{}{\rho}+\frac{1}{\rho^2}\secdiff{}{\phi}\right)\psi(\rho,\phi)
\\
&+\left[U(\rho,\phi)
  + \lambda
  g_4 \sum_{i<j}^{3}{\delta\left(\alpha
      \rho\left|\sin(\phi-\phi_{ij})\right|\right)}\right]\psi(\rho,\phi) \\
& = E_{EAA}\psi(\rho,\phi)
\end{split}
\end{equation}
Finally, we describe $\psi(\rho,\phi)$ as a sum over adiabatic channel
functions:
\begin{equation}
\label{psiexpansion}
\psi(\rho,\phi) = \sum_{n=0}^{\infty}{\rho^{-1/2}f_n(\rho)\chi_n(\rho;\phi)},
\end{equation}
where $\chi_n(\rho;\phi)$ satisfy the fixed-$\rho$ equation:
\begin{equation}
\label{fixedrhoeq}
\begin{split}
&\frac{-1}{2\mu_4\rho^2}\secdiff{\chi_n}{\phi}+U(\rho,\phi)\chi_n  \\
& + \lambda g_4 \sum_{i<j}^{3}{\delta\left(\alpha
      \rho\left|\sin(\phi-\phi_{ij})\right|\right)}\chi_n =
  U_n(\rho)\chi_n
\end{split}
\end{equation}
 Because we only
consider $\lambda=0$ and $\lambda\rightarrow\infty$, the
$\delta$-functions in Eq.~(\ref{fixedrhoeq}) result in simple boundary
conditions at $\phi=\pi/6$.  For arbitrary $\lambda$, one would
need to account for the $\lambda$-dependent derivative discontinuity
at $\phi=\pi/6$.  Inserting the expansion Eq.~(\ref{psiexpansion}) into
Eq.~(\ref{BOEquation}) results in a set of coupled equations in
$\rho$, which are conveniently written in matrix form as:
\begin{equation}
\label{hyperradialeqs}
\begin{split}
\frac{-1}{2\mu_4}&\left(\mathbf{1}\secdiff{}{\rho}+\mathbf{Q}(\rho)+2\mathbf{P}(\rho)\diff{}{\rho}\right)\vec{f}(\rho)\\
&+\mathbf{U}_{\text{eff}}(\rho)\vec{f}(\rho)=E_{EAA}\vec{f}(\rho)
\end{split}
\end{equation}
Here, $\mathbf{U}_{\text{eff}}$ is a diagonal matrix with elements $U_n(\rho)
- 1/8\mu_4\rho^2$, $P_{mn}(\rho)=\left\langle
  \chi_m\left|\diff{\chi_n}{\rho}\right.\right\rangle_\phi$ and $Q_{mn}(\rho)=\left\langle \chi_m
  \left|\secdiff{\chi_n}{\rho}\right.\right\rangle_\phi$.  When
$\mathbf{P}$ and $\mathbf{Q}$ are included in the solution to
Eq.~(\ref{hyperradialeqs}), and enough channels are retained for
numerical convergence, the accuracy of the four-body energy is (in
principle) limited only by the omission of first and second derivative
couplings,
$\left\langle \Phi_m \left|\vec{\nabla}\Phi_n\right.\right\rangle_z$
and $\left\langle \Phi_m
  \left|\nabla^2 \Phi_n \right.\right\rangle_z$, that arise from
generalizing Eq.~(\ref{BOfact}) to include a sum: $\Psi =
\sum_n{\Phi_n\psi_n}$.  Such a generalization is not possible for our
model without the introduction of a confining potential because
Eq.~(\ref{HHHLAdSE}) admits only one solution that vanishes as $|z|\rightarrow\infty$.

Identical particle symmetry of the heavy particles allows one to
restrict the domain of the four-body wavefunction to the region
$0<\phi<\pi/6$.  Thus, for a given
permutation of heavy particles, the locus of points describing the
coalescence of a heavy particle and a light particle --- i.e. when $z$
is equal to $z_i$ --- remain \emph{ordered} $z_1<z_2<z_3$ along the
$z$-coordinate.  Because
the ordering is independent of $\rho$ and $\phi$, the
solution to Eq.~(\ref{HHHLAdSE}) for all $\rho$ and all $\phi \in
[0,\pi/6]$ is straightforward.  

The boundary condition on $\chi(\rho;\phi)$ at $\phi=0$ is
determined by a combination of the parity operator,
$\hat{\Pi}\phi\rightarrow \phi+\pi$, and the 1-2
permutation operator, $\hat{P}_{12}\phi\rightarrow\pi-\phi$, by the rule: $\hat{P}_{12}\hat{\Pi}\phi\rightarrow-\phi$.
Considering positive parity, the boundary conditions on $\chi(\rho;\phi)$ for noninteracting
bosons are:
 $\left. \diff{\chi}{\phi}\right|_{\phi=0}=\left. \diff{\chi}{\phi}\right|_{\phi=\pi/6}=0$,
while for noninteracting fermions,
$\chi|_{\phi=0}=\chi|_{\phi=\pi/6}=0$.  Note that the boundary
conditions for noninteracting fermions of positive parity are
equivalent to those for bosons of negative parity, but $\lambda
\rightarrow \infty$.

\subsection{Numerical Solutions for the HHHL System}
In Appendix~\ref{app:tripledelta}, we calculate the
transcendental equation for the eigenvalue of a 1D Schr\"odinger
equation with three $\delta$-functions of arbitrary strength and
arbitrary placement.  The resulting Eq.~(\ref{tripledelta}) is
applied to the Eq.~(\ref{HHHLAdSE}) by letting
$g_a=g_b=g_c=2\mu_4g_4$, $\kappa^2 = -2\mu_4 U(\rho,\phi)>0$, and
$a=z_1$, $b=z_2$ and $c=z_3$.

\begin{figure}[!t]
  \begin{center}
    \leavevmode
    \includegraphics[width=3.5in,clip=true]{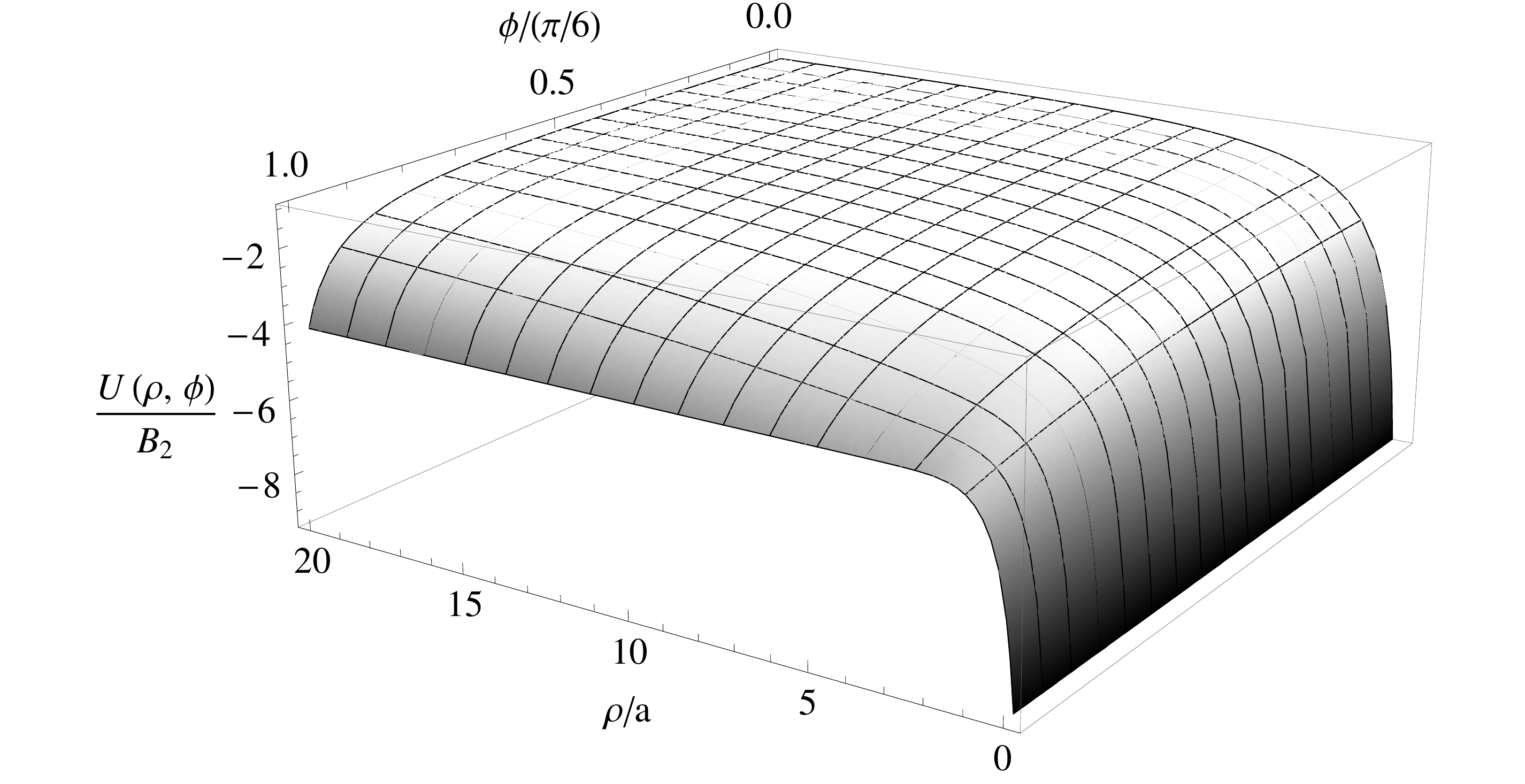}
    \caption{The Born-Oppenheimer surface for the HHHL system is
      shown.  $U(\rho,\phi$) is in units of $B_2$, and $\rho$ is in
      units of $a$.}
    \label{fig:HHHLSurface}
  \end{center}
\end{figure}

Figure.~\ref{fig:HHHLSurface} shows the potential energy surface
for the particular mass ratio $\beta^{-1}=22.08$ appropriate for an atomic mixture of
Li-Cs.  Potential surfaces like this one are calculated by solving
Eq.~(\ref{tripledelta}) on a nonlinear grid with typically $200 \times
400$ points in the $(\rho,\phi)$ plane.  The points are distributed so that more grid-points are concentrated in the vicinity of the well at $\rho=0$
and the valley near $\phi=\pi/6$.  Particular care must be taken to
describe the valley near $\phi=\pi/6$ accurately at large $\rho$, or else the
numerical solution to the fixed-$\rho$ adiabatic equation does not
reproduce the correct threshold behavior in any of the atom-trimer channels.
This is because the fixed-$\rho$ solutions as $\rho\rightarrow
\infty$ should approach the HHL bound-state energies from the
spectrum in Fig.~\ref{fig:HHLSpectrum} with the correct
$\rho$-dependence.  In particular, at large $\rho$ we find that $Q_{00}(\rho)/2\mu_4\rightarrow
-1/8\mu_4\rho^2$, and exactly cancels the $+1/8\mu_4\rho^2$ in the
$U_{0,\text{eff}}$.  That is, the effective potential with the
diagonal correction approaches a constant, and describes a 2-body
channel to which Eqs.~(\ref{matching}) and~(\ref{ere}) may be applied
with the replacements $\psi \rightarrow f$, $k_{AD}\rightarrow
k_{AT}=\sqrt{2\mu_{234,1}E_{\text{rel}}}$, $a_{AD}\rightarrow a_{AT}$,
and $x\rightarrow \rho$, along with the boundary condition $f(0)=0$.

Figures~\ref{fig:LiCsCurves}(a) and~\ref{fig:LiCsCurves}(c) show the
hyperradial potential curves $U_n(\rho)$ obtained by solving the
fixed-$\rho$ Eq.~(\ref{fixedrhoeq}) using the potential energy surface
shown in Fig.~\ref{fig:HHHLSurface}.  Note that at large $\rho$, the
lowest potential curves converge to the appropriate HHL bound state
energy shown as red dashed lines in Fig.~\ref{fig:LiCsCurves}(b)
and~\ref{fig:LiCsCurves}(d), as appropriate for an atom-trimer
channel.  The red dashed lines in Fig.~\ref{fig:LiCsCurves}(a) and (c)
indicate HHHL bound states obtained by solving
Eq.~(\ref{hyperradialeqs}) with 10 coupled channels. Typically,
calculations with only the lowest channel (but including the diagonal
correction) give bound-state energies converged to four or five
digits.  The error incurred by ignoring excited hyperradial potential
curves is expected to be negligible compared to making the EAA
in the calculation of the surface $U(\rho,\phi)$. 

\begin{center}
  \begin{figure}[!t]
    \includegraphics[width=3.5in,clip=true]{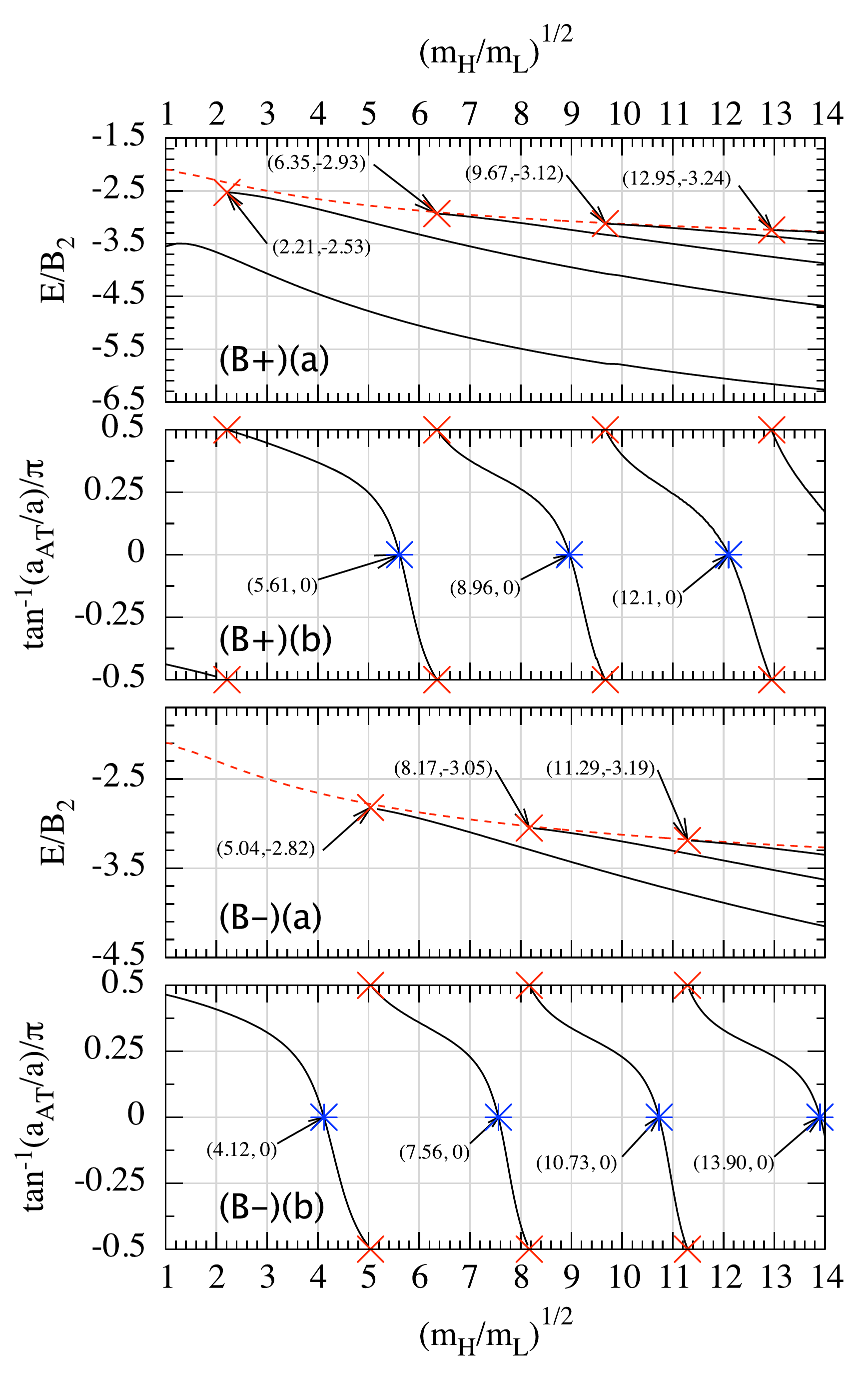}
    \caption{Graphs (B+)(a) and (B-)(a) show the HHHL spectrum for bosonic heavy
      particles of positive and negative parity, respectively.  The
      red crosses indicate the appearance of a new HHHL bound state
      and $a_{AT}\rightarrow \infty$.  The red dashed curve is the
      lowest (solid line) bosonic HHL bound state from
      Fig.~\ref{fig:HHLSpectrum}(a).  Graphs (B+)(b) and (B-)(b) show the arctangent
      of the atom-trimer scattering length for bosonic heavy particles
      of positive and negative parity, respectively.  The blue stars indicate
    $a_{AT}\rightarrow 0$.  The specific ordinates of the 
    crosses and stars are labeled.}
    \label{fig:ATScatLenBosons}
  \end{figure}
\end{center}

\begin{center}
  \begin{figure}[!t]
    \includegraphics[width=3.5in,clip=true]{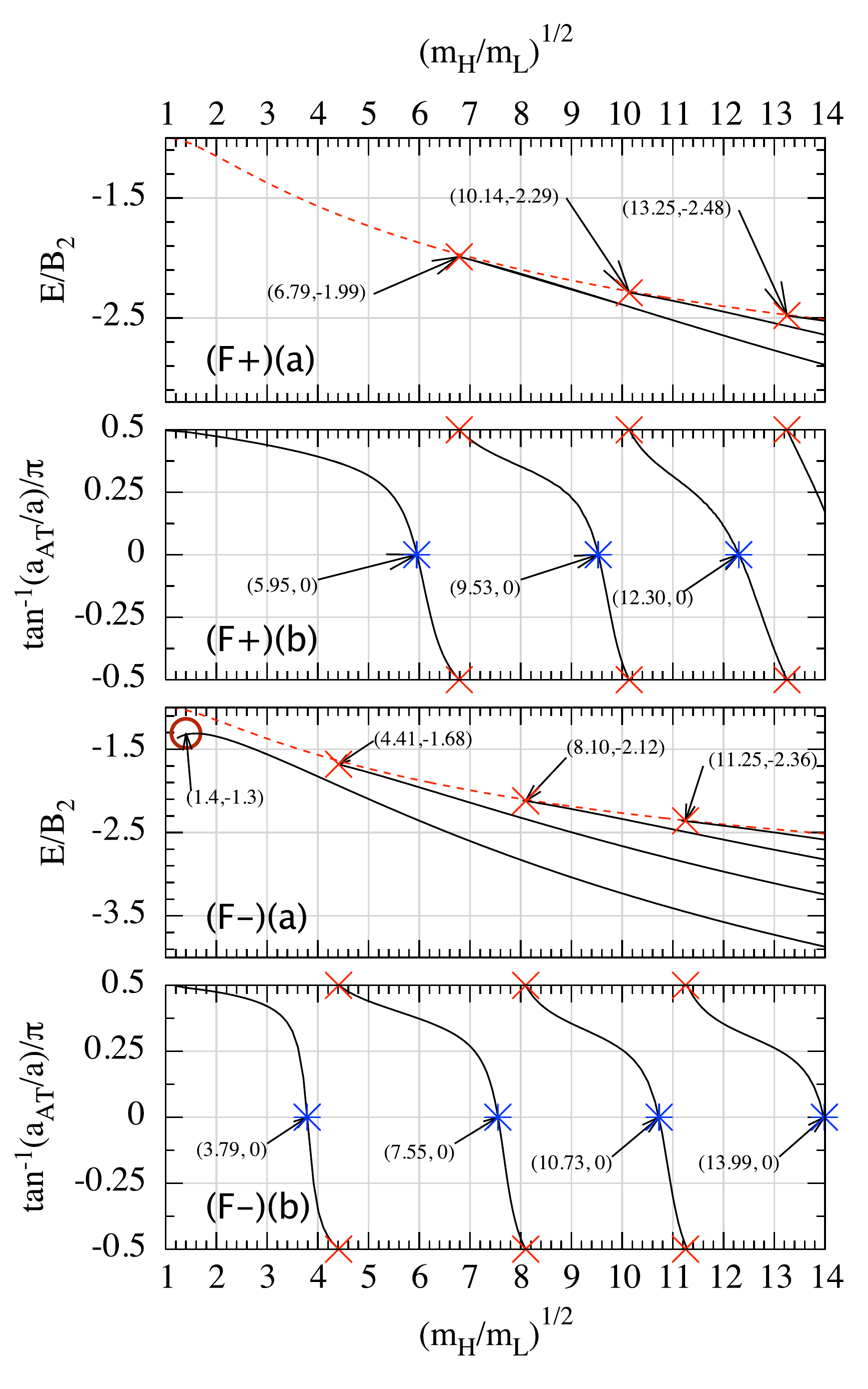}
    \caption{Graphs (F+)(a) and (F-)(a) show the HHHL spectrum for fermionic heavy
      particles of positive and negative parity, respectively.  The
      red crosses indicate the appearance of a new HHHL bound state
      and $a_{AT}\rightarrow \infty$.  The red dashed curve is the
      lowest (solid line) fermionic HHL bound state from
      Fig.~\ref{fig:HHLSpectrum}(a).  Graphs (F+)(b) and (F-)(b) show the arctangent
      of the atom-trimer scattering length for fermionic heavy particles
      of positive and negative parity, respectively.  The blue stars indicate
    $a_{AT}\rightarrow 0$.  The specific ordinates of the 
    crosses and stars are labeled.}
    \label{fig:ATScatLenFermions}
  \end{figure}
\end{center}

In Figures~\ref{fig:ATScatLenBosons} and~\ref{fig:ATScatLenFermions},
we show the spectrum and atom-trimer scattering lengths of the HHHL system with noninteracting bosonic and fermionic heavy atoms,
respectively.  We show both positive (B+, F+), and negative (B-, F-) parity
cases for each identical particle symmetry.  The HHL ground-state energies for each symmetry from
Fig.~\ref{fig:HHLSpectrum}(a) are replotted here as dashed-red
curves.  Again, mass ratios at which a new
tetramer state appears (and $|a_{AT}|\rightarrow\infty$) are marked by
red crosses, while zeroes of $a_{AT}$ are indicated by blue stars.
The particular numerical values for the coordinates
$(\beta^{-1/2},E/B_2)$ are also marked.  As the mass-ratio
$\beta^{-1}$ increases, four-body bound states
enter at lower energies than one would expect from the three-body
calculation (i.e. the dashed-red curve).  This discrepancy in the threshold energy
is attributed to the fact that the EAA underestimates
the potential surface in Fig.~\ref{fig:HHHLSurface} by neglecting the
positive nonadiabatic correction
$-\left\langle\Phi\right|\left.\nabla^2\Phi\right\rangle_z/2\mu_4$,
while the corresponding correction is included at the three-body level
in Fig.~\ref{fig:HHLSpectrum}. 

While we perform a multichannel calculation of the spectrum, we find
that it is sufficient to use a simple single-channel calculation for
$a_{AT}$.  Indeed, comparing the
critical mass ratios for which $a_{AT}\rightarrow \infty$, and a
tetramer state lies at threshold, we find
good agreement between the two calculations.  This can be readily
observed by comparing the positions of the red crosses in graphs (a) and (b) of
Fig.~\ref{fig:ATScatLenBosons}, and similarly in Fig.~\ref{fig:ATScatLenFermions}.

At $\beta = 1$, we find that noninteracting bosons of positive parity admit an HHHL
bound state with $E_{4,\text{EAA}} \approx -3.55$.  The second tetramer state appears at
$\beta^{-1} \approx 4.88$, and the third at $\beta^{-1}\approx 40.3$.
For Li-Cs mixtures, one might expect two universal tetramer states.
Negative parity bosons are less likely to bind than those with
positive parity.  The first tetramer
state appears at $\beta^{-1}\approx 25.4$, and the second at
$\beta^{-1}\approx 66.7$.

For fermionic particles, negative parity tetramers are more likely to
bind than those with positive parity.  The precise value of the
critical mass ratio is difficult to pin down within the
Born-Oppenheimer approximation at these small mass ratios.  The
difficulty is magnified because even in the UAA three-body
calculations of Section~\ref{HHLSector}, the trimer state doesn't
appear until $\beta^{-1}\approx 1.17$.  Because the four-body
calculation doesn't include the positive nonadiabatic correction to the potential energy
surface shown in Fig.~\ref{fig:HHHLSurface}, One expects the tetramer
bound state to appear below the atom-trimer threshold prematurely.  In the four-body
calculation, the energy of the atom trimer threshold itself increases
slightly with $\beta^{-1/2}$, and the tetramer energy tracks along
with it until about $\beta^{-1}\approx 2.6$.  The increasing threshold
energy is undoubtedly an artifact of the approximation at the
four-body level since it is absent in the more accurate three-body
calculations.  We can nonetheless estimate the critical mass ratio as
$\beta^{-1}\approx 2.0\pm 0.6$, indicated by a red circle in
Fig.~\ref{fig:ATScatLenFermions}(F-)(a).  The second negative parity
fermionic state appears at $\beta^{-1}=19.4$. 

In 3D, Blume~\cite{Blume2012PRL} found that a universal tetramer exists for
fermionic particles above a mass ratio of $\beta^{-1}\approx
9.5$.  In 2D, Levinsen and Parish~\cite{Levinsen2013PRL} found a
critical mass ratio of $\beta^{-1}\approx 5.0$.  It is interesting to
speculate whether these states are continuously connected to each
other, and to the universal state that appears in these calculations
at $\beta^{-1}\approx 2.0$ for negative parity fermions.

\section{Summary and Outlook}
\label{summary}

We have calculated three-body and four-body spectra, as well as the
atom-dimer and atom-trimer scattering lengths for two-component systems with one
light particle, as a function of the mass ratio.  Heavy
particles are assumed to be noninteracting, and the four-particle
system is assumed to be in free-space.  Both bosonic and fermionic
heavy particles are treated.  For the HHL system, the
Born-Oppenheimer method gives good quantitative agreement with the
hyperspherical calculations of~\cite{Kartavtsev2009JETP}. For the HHHL
system, the potential energy surface governing the heavy-particle
dynamics is calculated in the ``extreme adiabatic approximation''.
That surface is then used to calculate hyperradial potential curves
and couplings.  The values for the resulting atom-trimer thresholds
converge to the appropriate three-body bound state energies,
lending some confidence to the HHHL calculation.

Let us now discuss possible extensions of this work.  Note that we
have scaled away the only length scale, $a$, that appears in our
model.  There are two immediate generalizations that expand the
parameter space considerably.  

First, there is the generalization to arbitrary H-H interactions, which
introduces the H-H scattering length $a_{HH}$.  Such an extension was
already treated at the three-body level in~\cite{Kartavtsev2009JETP},
but no such four-body calculations have appeared in the literature.
In a hyperspherical calculation, the additional derivative
discontinuity in the angular wavefunction is treated analytically, and
the hyperradial potential curves are calculated as the solution to a
single transcendental equation~\cite{Mehta2007PRA,Kartavtsev2009JETP}.
The HHHL Born-Oppenheimer calculation for bosons can be extended to
arbitrary $\lambda$ by choosing a b-spline basis set that satisfies
the boundary condition,
\begin{equation}
\lim_{\epsilon \rightarrow
  0}{\left.\frac{1}{\chi}\diff{\chi}{\phi}\right|_{\pi/6-\epsilon}}=\frac{\rho
    \lambda (1+\beta)}{\sqrt{2}\beta}\left(\frac{\beta}{3+\beta}\right)^{5/6}
\end{equation}
With this generalization, one can smoothly transition between
the energies shown in Fig.~\ref{fig:ATScatLenBosons} and
Fig.~\ref{fig:ATScatLenFermions}, passing from noninteracting bosons
to the fermionized limit.

The bound-state calculation can be extended by the introduction of a
harmonic trapping potential, which separates into relative and
center-of-mass parts under the transformation to Jacobi coordinates.
This extension would establish a connection with several papers that
have appeared recently, treating equal-mass two-component
systems~\cite{Gharashi2013PRL,Sowinski2013PRA,Lindgren2013Arxiv}.  The
addition of a trapping potential would introduce excited potential
energy surfaces and the possibility of interesting physics beyond the
Born-Oppenheimer approximation.

Here, we have only considered the ``3+1'' branch (i.e. the HHHL
system) of the few-component problem.  It may be that other branches
can be treated by similar methods.  For example, for the 2+2 (HHLL)
problem, integrating out the light atoms would result in a 1D
heavy-heavy potential, but the adiabatic equation is a 2D partial
differential equation, instead of a 1D equation like
Eq.~(\ref{HHHLAdSE}).

Finally, it is worth emphasizing that ultimately a fully 3D solution
to the few-body problem with finite-range interactions is
needed in order to understand the physics of quasi-1D systems.
A hyperspherical solution to the few-body problem in quasi-1D
for finite-range interactions remains a significant challenge,
although recent advances in the Correlated Gaussian Hyperspherical
method~\cite{Daily2014PRA} may make these calculations possible.

\begin{acknowledgments}
I would like to thank Brett Esry, Chris Greene and Jose D'Incao for early
discussions related to this topic.  Thanks also to Jesper Levinsen for
a series of helpful correspondences.  
\end{acknowledgments}

\appendix
\section{Triple $\delta$-function problem}
\label{app:tripledelta}
Here, we give the solution for the eigenvalue to the following Schr\"odinger equation:
\begin{equation}
\begin{split}
&\left[-\secdiff{}{z}+g_a \delta(z-a)+g_b \delta(z-b)+g_c
\delta(z-c)\right]\Phi(z) \\
& =-\kappa^2\Phi(z)
\end{split}
\end{equation}
We assume that the positions of the $\delta$-functions are ordered as $a<b<c$, but no other assumptions regarding their placement are made.  In particular, the Hamiltonian is not assumed to commute with the parity operator.  The (unnormalized) solution satisfying asymptotic boundary condition, $\Phi(|z|\rightarrow \infty)=0$, is elementary:
\begin{equation}
\Phi(z)=
\begin{cases}
 \Phi_I&=A e^{\kappa  z} \;\;  z<a \\
 \Phi _{\text{II}}&=Be^{-\kappa  z} +C e^{\kappa  z} \;\; a<z<b \\
 \Phi _{\text{III}}&=De^{-\kappa  z} +E e^{\kappa  z} \;\;b<z<c \\
 \Phi _{\text{IV}}&=F e^{-\kappa  z}\;\; c<z 
\end{cases}
\end{equation}
Matching the solutions and enforcing the derivative discontinuities at
$z=a$, $z=b$, and $z=c$ yields, after considerable algebra:
\begin{align}
\label{tripledelta}
&g_a g_c (g_b-2 \kappa ) e^{2 \kappa  (a+b)}-g_a g_b (g_c+2 \kappa ) e^{2 \kappa  (a+c)} \notag\\
&+(g_a+2 \kappa ) (g_b+2 \kappa ) (g_c+2 \kappa ) e^{2 \kappa  (b+c)} \notag\\
&-g_b g_c e^{4 b \kappa } (g_a+2 \kappa ) = 0
\end{align}
For the special case of a quadrupolar potential ($a=-c$, $b=0$ and
$g_a=g_c=-g_b/2$), Eq.~\ref{tripledelta} reduces to the result found
recently by Patil~\cite{Patil2009EJP}.

\bibliography{../../../../AllRefs.bib}

\begin{thebibliography}{49}%
\makeatletter
\providecommand \@ifxundefined [1]{%
 \@ifx{#1\undefined}
}%
\providecommand \@ifnum [1]{%
 \ifnum #1\expandafter \@firstoftwo
 \else \expandafter \@secondoftwo
 \fi
}%
\providecommand \@ifx [1]{%
 \ifx #1\expandafter \@firstoftwo
 \else \expandafter \@secondoftwo
 \fi
}%
\providecommand \natexlab [1]{#1}%
\providecommand \enquote  [1]{``#1''}%
\providecommand \bibnamefont  [1]{#1}%
\providecommand \bibfnamefont [1]{#1}%
\providecommand \citenamefont [1]{#1}%
\providecommand \href@noop [0]{\@secondoftwo}%
\providecommand \href [0]{\begingroup \@sanitize@url \@href}%
\providecommand \@href[1]{\@@startlink{#1}\@@href}%
\providecommand \@@href[1]{\endgroup#1\@@endlink}%
\providecommand \@sanitize@url [0]{\catcode `\\12\catcode `\$12\catcode
  `\&12\catcode `\#12\catcode `\^12\catcode `\_12\catcode `\%12\relax}%
\providecommand \@@startlink[1]{}%
\providecommand \@@endlink[0]{}%
\providecommand \url  [0]{\begingroup\@sanitize@url \@url }%
\providecommand \@url [1]{\endgroup\@href {#1}{\urlprefix }}%
\providecommand \urlprefix  [0]{URL }%
\providecommand \Eprint [0]{\href }%
\providecommand \doibase [0]{http://dx.doi.org/}%
\providecommand \selectlanguage [0]{\@gobble}%
\providecommand \bibinfo  [0]{\@secondoftwo}%
\providecommand \bibfield  [0]{\@secondoftwo}%
\providecommand \translation [1]{[#1]}%
\providecommand \BibitemOpen [0]{}%
\providecommand \bibitemStop [0]{}%
\providecommand \bibitemNoStop [0]{.\EOS\space}%
\providecommand \EOS [0]{\spacefactor3000\relax}%
\providecommand \BibitemShut  [1]{\csname bibitem#1\endcsname}%
\let\auto@bib@innerbib\@empty
\bibitem [{\citenamefont {Greiner}\ \emph {et~al.}(2001)\citenamefont
  {Greiner}, \citenamefont {Bloch}, \citenamefont {Mandel}, \citenamefont
  {H{\"a}nsch},\ and\ \citenamefont {Esslinger}}]{Greiner2001PRL}%
  \BibitemOpen
  \bibfield  {author} {\bibinfo {author} {\bibfnamefont {M.}~\bibnamefont
  {Greiner}}, \bibinfo {author} {\bibfnamefont {I.}~\bibnamefont {Bloch}},
  \bibinfo {author} {\bibfnamefont {O.}~\bibnamefont {Mandel}}, \bibinfo
  {author} {\bibfnamefont {T.~W.}\ \bibnamefont {H{\"a}nsch}}, \ and\ \bibinfo
  {author} {\bibfnamefont {T.}~\bibnamefont {Esslinger}},\ }\href@noop {}
  {\bibfield  {journal} {\bibinfo  {journal} {Phys. Rev. Lett.}\ }\textbf
  {\bibinfo {volume} {87}},\ \bibinfo {pages} {160405} (\bibinfo {year}
  {2001})}\BibitemShut {NoStop}%
\bibitem [{\citenamefont {G\"orlitz}\ \emph {et~al.}(2001)\citenamefont
  {G\"orlitz}, \citenamefont {Vogels}, \citenamefont {Leanhardt}, \citenamefont
  {Raman}, \citenamefont {Gustavson}, \citenamefont {Abo-Shaeer}, \citenamefont
  {Chikkatur}, \citenamefont {Gupta}, \citenamefont {Inouye}, \citenamefont
  {Rosenband},\ and\ \citenamefont {Ketterle}}]{Gorlitz2001PRL}%
  \BibitemOpen
  \bibfield  {author} {\bibinfo {author} {\bibfnamefont {A.}~\bibnamefont
  {G\"orlitz}}, \bibinfo {author} {\bibfnamefont {J.~M.}\ \bibnamefont
  {Vogels}}, \bibinfo {author} {\bibfnamefont {A.~E.}\ \bibnamefont
  {Leanhardt}}, \bibinfo {author} {\bibfnamefont {C.}~\bibnamefont {Raman}},
  \bibinfo {author} {\bibfnamefont {T.~L.}\ \bibnamefont {Gustavson}}, \bibinfo
  {author} {\bibfnamefont {J.~R.}\ \bibnamefont {Abo-Shaeer}}, \bibinfo
  {author} {\bibfnamefont {A.~P.}\ \bibnamefont {Chikkatur}}, \bibinfo {author}
  {\bibfnamefont {S.}~\bibnamefont {Gupta}}, \bibinfo {author} {\bibfnamefont
  {S.}~\bibnamefont {Inouye}}, \bibinfo {author} {\bibfnamefont
  {T.}~\bibnamefont {Rosenband}}, \ and\ \bibinfo {author} {\bibfnamefont
  {W.}~\bibnamefont {Ketterle}},\ }\href@noop {} {\bibfield  {journal}
  {\bibinfo  {journal} {Phys. Rev. Lett.}\ }\textbf {\bibinfo {volume} {87}},\
  \bibinfo {pages} {130402} (\bibinfo {year} {2001})}\BibitemShut {NoStop}%
\bibitem [{\citenamefont {{B.~Laburthe Tolra and K.~M.~O'Hara and J.~H.~Huckans
  and W.~D.~Phillips and S.~L.~Rolston and J.~V.~Porto}}(2004)}]{Tolra2004PRL}%
  \BibitemOpen
  \bibfield  {author} {\bibinfo {author} {\bibnamefont {{B.~Laburthe Tolra and
  K.~M.~O'Hara and J.~H.~Huckans and W.~D.~Phillips and S.~L.~Rolston and
  J.~V.~Porto}}},\ }\href@noop {} {\bibfield  {journal} {\bibinfo  {journal}
  {Phys. Rev. Lett.}\ }\textbf {\bibinfo {volume} {92}},\ \bibinfo {pages}
  {190401} (\bibinfo {year} {2004})}\BibitemShut {NoStop}%
\bibitem [{\citenamefont {Kinoshita}\ \emph {et~al.}(2004)\citenamefont
  {Kinoshita}, \citenamefont {Wenger},\ and\ \citenamefont
  {Weiss}}]{Kinoshita2004Science}%
  \BibitemOpen
  \bibfield  {author} {\bibinfo {author} {\bibfnamefont {T.}~\bibnamefont
  {Kinoshita}}, \bibinfo {author} {\bibfnamefont {T.}~\bibnamefont {Wenger}}, \
  and\ \bibinfo {author} {\bibfnamefont {D.~S.}\ \bibnamefont {Weiss}},\
  }\href@noop {} {\bibfield  {journal} {\bibinfo  {journal} {Science}\ }\textbf
  {\bibinfo {volume} {305}},\ \bibinfo {pages} {1125} (\bibinfo {year}
  {2004})}\BibitemShut {NoStop}%
\bibitem [{\citenamefont {Kinoshita}\ \emph {et~al.}(2006)\citenamefont
  {Kinoshita}, \citenamefont {Wenger},\ and\ \citenamefont
  {Weiss}}]{Kinoshita2006Nature}%
  \BibitemOpen
  \bibfield  {author} {\bibinfo {author} {\bibfnamefont {T.}~\bibnamefont
  {Kinoshita}}, \bibinfo {author} {\bibfnamefont {T.}~\bibnamefont {Wenger}}, \
  and\ \bibinfo {author} {\bibfnamefont {D.~S.}\ \bibnamefont {Weiss}},\
  }\href@noop {} {\bibfield  {journal} {\bibinfo  {journal} {Nature}\ }\textbf
  {\bibinfo {volume} {440}},\ \bibinfo {pages} {900} (\bibinfo {year}
  {2006})}\BibitemShut {NoStop}%
\bibitem [{\citenamefont {Leanhardt}\ \emph {et~al.}(2002)\citenamefont
  {Leanhardt}, \citenamefont {Chikkatur}, \citenamefont {Kielpinski},
  \citenamefont {Shin}, \citenamefont {Gustavson}, \citenamefont {Ketterle},\
  and\ \citenamefont {Pritchard}}]{Leanhardt2002PRL}%
  \BibitemOpen
  \bibfield  {author} {\bibinfo {author} {\bibfnamefont {A.~E.}\ \bibnamefont
  {Leanhardt}}, \bibinfo {author} {\bibfnamefont {A.~P.}\ \bibnamefont
  {Chikkatur}}, \bibinfo {author} {\bibfnamefont {D.}~\bibnamefont
  {Kielpinski}}, \bibinfo {author} {\bibfnamefont {Y.}~\bibnamefont {Shin}},
  \bibinfo {author} {\bibfnamefont {T.~L.}\ \bibnamefont {Gustavson}}, \bibinfo
  {author} {\bibfnamefont {W.}~\bibnamefont {Ketterle}}, \ and\ \bibinfo
  {author} {\bibfnamefont {D.~E.}\ \bibnamefont {Pritchard}},\ }\href {\doibase
  10.1103/PhysRevLett.89.040401} {\bibfield  {journal} {\bibinfo  {journal}
  {Phys. Rev. Lett.}\ }\textbf {\bibinfo {volume} {89}},\ \bibinfo {pages}
  {040401} (\bibinfo {year} {2002})}\BibitemShut {NoStop}%
\bibitem [{\citenamefont {Fried}\ \emph {et~al.}(1998)\citenamefont {Fried},
  \citenamefont {Killian}, \citenamefont {Willmann}, \citenamefont {Landhuis},
  \citenamefont {Moss}, \citenamefont {Kleppner},\ and\ \citenamefont
  {Greytak}}]{Fried1998PRL}%
  \BibitemOpen
  \bibfield  {author} {\bibinfo {author} {\bibfnamefont {D.~G.}\ \bibnamefont
  {Fried}}, \bibinfo {author} {\bibfnamefont {T.~C.}\ \bibnamefont {Killian}},
  \bibinfo {author} {\bibfnamefont {L.}~\bibnamefont {Willmann}}, \bibinfo
  {author} {\bibfnamefont {D.}~\bibnamefont {Landhuis}}, \bibinfo {author}
  {\bibfnamefont {S.~C.}\ \bibnamefont {Moss}}, \bibinfo {author}
  {\bibfnamefont {D.}~\bibnamefont {Kleppner}}, \ and\ \bibinfo {author}
  {\bibfnamefont {T.~J.}\ \bibnamefont {Greytak}},\ }\href@noop {} {\bibfield
  {journal} {\bibinfo  {journal} {Phys. Rev. Lett.}\ }\textbf {\bibinfo
  {volume} {81}},\ \bibinfo {pages} {3811} (\bibinfo {year}
  {1998})}\BibitemShut {NoStop}%
\bibitem [{\citenamefont {Parker}\ \emph {et~al.}(2012)\citenamefont {Parker},
  \citenamefont {Dietrich}, \citenamefont {Bailey}, \citenamefont {Greene},
  \citenamefont {Holt}, \citenamefont {Kalita}, \citenamefont {Korsch},
  \citenamefont {Lu}, \citenamefont {Mueller}, \citenamefont {O'Connor},
  \citenamefont {Singh}, \citenamefont {Sulai},\ and\ \citenamefont
  {Trimble}}]{Parker2012PRC}%
  \BibitemOpen
  \bibfield  {author} {\bibinfo {author} {\bibfnamefont {R.~H.}\ \bibnamefont
  {Parker}}, \bibinfo {author} {\bibfnamefont {M.~R.}\ \bibnamefont
  {Dietrich}}, \bibinfo {author} {\bibfnamefont {K.}~\bibnamefont {Bailey}},
  \bibinfo {author} {\bibfnamefont {J.~P.}\ \bibnamefont {Greene}}, \bibinfo
  {author} {\bibfnamefont {R.~J.}\ \bibnamefont {Holt}}, \bibinfo {author}
  {\bibfnamefont {M.~R.}\ \bibnamefont {Kalita}}, \bibinfo {author}
  {\bibfnamefont {W.}~\bibnamefont {Korsch}}, \bibinfo {author} {\bibfnamefont
  {Z.-T.}\ \bibnamefont {Lu}}, \bibinfo {author} {\bibfnamefont
  {P.}~\bibnamefont {Mueller}}, \bibinfo {author} {\bibfnamefont {T.~P.}\
  \bibnamefont {O'Connor}}, \bibinfo {author} {\bibfnamefont {J.}~\bibnamefont
  {Singh}}, \bibinfo {author} {\bibfnamefont {I.~A.}\ \bibnamefont {Sulai}}, \
  and\ \bibinfo {author} {\bibfnamefont {W.~L.}\ \bibnamefont {Trimble}},\
  }\href {\doibase 10.1103/PhysRevC.86.065503} {\bibfield  {journal} {\bibinfo
  {journal} {Phys. Rev. C}\ }\textbf {\bibinfo {volume} {86}},\ \bibinfo
  {pages} {065503} (\bibinfo {year} {2012})}\BibitemShut {NoStop}%
\bibitem [{\citenamefont {Ni}\ \emph {et~al.}(2008)\citenamefont {Ni},
  \citenamefont {Ospelkaus}, \citenamefont {De~Miranda}, \citenamefont {Pe'er},
  \citenamefont {Neyenhuis}, \citenamefont {Zirbel}, \citenamefont
  {Kotochigova}, \citenamefont {Julienne}, \citenamefont {Jin},\ and\
  \citenamefont {Ye}}]{Ni2008Science}%
  \BibitemOpen
  \bibfield  {author} {\bibinfo {author} {\bibfnamefont {K.-K.}\ \bibnamefont
  {Ni}}, \bibinfo {author} {\bibfnamefont {S.}~\bibnamefont {Ospelkaus}},
  \bibinfo {author} {\bibfnamefont {M.}~\bibnamefont {De~Miranda}}, \bibinfo
  {author} {\bibfnamefont {A.}~\bibnamefont {Pe'er}}, \bibinfo {author}
  {\bibfnamefont {B.}~\bibnamefont {Neyenhuis}}, \bibinfo {author}
  {\bibfnamefont {J.}~\bibnamefont {Zirbel}}, \bibinfo {author} {\bibfnamefont
  {S.}~\bibnamefont {Kotochigova}}, \bibinfo {author} {\bibfnamefont
  {P.}~\bibnamefont {Julienne}}, \bibinfo {author} {\bibfnamefont
  {D.}~\bibnamefont {Jin}}, \ and\ \bibinfo {author} {\bibfnamefont
  {J.}~\bibnamefont {Ye}},\ }\href@noop {} {\bibfield  {journal} {\bibinfo
  {journal} {Science}\ }\textbf {\bibinfo {volume} {322}},\ \bibinfo {pages}
  {231} (\bibinfo {year} {2008})}\BibitemShut {NoStop}%
\bibitem [{\citenamefont {Barontini}\ \emph {et~al.}(2009)\citenamefont
  {Barontini}, \citenamefont {Weber}, \citenamefont {Rabatti}, \citenamefont
  {Catani}, \citenamefont {Thalhammer}, \citenamefont {Inguscio},\ and\
  \citenamefont {Minardi}}]{Barontini2009PRL}%
  \BibitemOpen
  \bibfield  {author} {\bibinfo {author} {\bibfnamefont {G.}~\bibnamefont
  {Barontini}}, \bibinfo {author} {\bibfnamefont {C.}~\bibnamefont {Weber}},
  \bibinfo {author} {\bibfnamefont {F.}~\bibnamefont {Rabatti}}, \bibinfo
  {author} {\bibfnamefont {J.}~\bibnamefont {Catani}}, \bibinfo {author}
  {\bibfnamefont {G.}~\bibnamefont {Thalhammer}}, \bibinfo {author}
  {\bibfnamefont {M.}~\bibnamefont {Inguscio}}, \ and\ \bibinfo {author}
  {\bibfnamefont {F.}~\bibnamefont {Minardi}},\ }\href {\doibase
  10.1103/PhysRevLett.103.043201} {\bibfield  {journal} {\bibinfo  {journal}
  {Phys. Rev. Lett.}\ }\textbf {\bibinfo {volume} {103}},\ \bibinfo {pages}
  {043201} (\bibinfo {year} {2009})}\BibitemShut {NoStop}%
\bibitem [{\citenamefont {Hara}\ \emph {et~al.}(2011)\citenamefont {Hara},
  \citenamefont {Takasu}, \citenamefont {Yamaoka}, \citenamefont {Doyle},\ and\
  \citenamefont {Takahashi}}]{Hara2011PRL}%
  \BibitemOpen
  \bibfield  {author} {\bibinfo {author} {\bibfnamefont {H.}~\bibnamefont
  {Hara}}, \bibinfo {author} {\bibfnamefont {Y.}~\bibnamefont {Takasu}},
  \bibinfo {author} {\bibfnamefont {Y.}~\bibnamefont {Yamaoka}}, \bibinfo
  {author} {\bibfnamefont {J.~M.}\ \bibnamefont {Doyle}}, \ and\ \bibinfo
  {author} {\bibfnamefont {Y.}~\bibnamefont {Takahashi}},\ }\href@noop {}
  {\bibfield  {journal} {\bibinfo  {journal} {Phys. Rev. Lett.}\ }\textbf
  {\bibinfo {volume} {106}},\ \bibinfo {pages} {205304} (\bibinfo {year}
  {2011})}\BibitemShut {NoStop}%
\bibitem [{\citenamefont {Lieb}\ and\ \citenamefont
  {Liniger}(1963)}]{LiebLiniger1963PhysRev}%
  \BibitemOpen
  \bibfield  {author} {\bibinfo {author} {\bibfnamefont {E.}~\bibnamefont
  {Lieb}}\ and\ \bibinfo {author} {\bibfnamefont {W.}~\bibnamefont {Liniger}},\
  }\href@noop {} {\bibfield  {journal} {\bibinfo  {journal} {Phys. Rev.}\
  }\textbf {\bibinfo {volume} {130}},\ \bibinfo {pages} {1605} (\bibinfo {year}
  {1963})}\BibitemShut {NoStop}%
\bibitem [{\citenamefont {{J.~B.~McGuire}}(1964)}]{McGuire1964JMP}%
  \BibitemOpen
  \bibfield  {author} {\bibinfo {author} {\bibnamefont {{J.~B.~McGuire}}},\
  }\href@noop {} {\bibfield  {journal} {\bibinfo  {journal} {J. Math. Phys.}\
  }\textbf {\bibinfo {volume} {5}},\ \bibinfo {pages} {622} (\bibinfo {year}
  {1964})}\BibitemShut {NoStop}%
\bibitem [{\citenamefont {Yang}(1967)}]{Yang1967PRL}%
  \BibitemOpen
  \bibfield  {author} {\bibinfo {author} {\bibfnamefont {C.~N.}\ \bibnamefont
  {Yang}},\ }\href@noop {} {\bibfield  {journal} {\bibinfo  {journal} {Phys.
  Rev. Lett.}\ }\textbf {\bibinfo {volume} {19}},\ \bibinfo {pages} {1312}
  (\bibinfo {year} {1967})}\BibitemShut {NoStop}%
\bibitem [{\citenamefont {Yang}\ and\ \citenamefont
  {Yang}(1969)}]{YangYang1969JMP}%
  \BibitemOpen
  \bibfield  {author} {\bibinfo {author} {\bibfnamefont {C.}~\bibnamefont
  {Yang}}\ and\ \bibinfo {author} {\bibfnamefont {C.}~\bibnamefont {Yang}},\
  }\href@noop {} {\bibfield  {journal} {\bibinfo  {journal} {J. Math. Phys.}\
  }\textbf {\bibinfo {volume} {10}},\ \bibinfo {pages} {1115} (\bibinfo {year}
  {1969})}\BibitemShut {NoStop}%
\bibitem [{\citenamefont {Dodd}(1970)}]{Dodd1970JMP}%
  \BibitemOpen
  \bibfield  {author} {\bibinfo {author} {\bibfnamefont {L.}~\bibnamefont
  {Dodd}},\ }\href@noop {} {\bibfield  {journal} {\bibinfo  {journal} {J. Math.
  Phys.}\ }\textbf {\bibinfo {volume} {11}},\ \bibinfo {pages} {207} (\bibinfo
  {year} {1970})}\BibitemShut {NoStop}%
\bibitem [{\citenamefont {{H.~B.~Thacker}}(1974)}]{Thacker1974PRD}%
  \BibitemOpen
  \bibfield  {author} {\bibinfo {author} {\bibnamefont {{H.~B.~Thacker}}},\
  }\href@noop {} {\bibfield  {journal} {\bibinfo  {journal} {Phys.~Rev.~D}\
  }\textbf {\bibinfo {volume} {11}},\ \bibinfo {pages} {838} (\bibinfo {year}
  {1974})}\BibitemShut {NoStop}%
\bibitem [{\citenamefont {Gibson}\ \emph {et~al.}(1987)\citenamefont {Gibson},
  \citenamefont {Larsen},\ and\ \citenamefont {Popiel}}]{Gibson1987PRA}%
  \BibitemOpen
  \bibfield  {author} {\bibinfo {author} {\bibfnamefont {W.~G.}\ \bibnamefont
  {Gibson}}, \bibinfo {author} {\bibfnamefont {S.~Y.}\ \bibnamefont {Larsen}},
  \ and\ \bibinfo {author} {\bibfnamefont {J.}~\bibnamefont {Popiel}},\
  }\href@noop {} {\bibfield  {journal} {\bibinfo  {journal} {Phys. Rev. A}\
  }\textbf {\bibinfo {volume} {35}},\ \bibinfo {pages} {4919} (\bibinfo {year}
  {1987})}\BibitemShut {NoStop}%
\bibitem [{\citenamefont {{A.~Amaya-Tapia and S.Y.~Larsen and
  J.~Popiel}}(1997)}]{Amaya-Tapia1997FBS}%
  \BibitemOpen
  \bibfield  {author} {\bibinfo {author} {\bibnamefont {{A.~Amaya-Tapia and
  S.Y.~Larsen and J.~Popiel}}},\ }\href@noop {} {\bibfield  {journal} {\bibinfo
   {journal} {Few-Body Syst.}\ }\textbf {\bibinfo {volume} {23}},\ \bibinfo
  {pages} {87} (\bibinfo {year} {1997})}\BibitemShut {NoStop}%
\bibitem [{\citenamefont {Mehta}\ and\ \citenamefont
  {Shepard}(2005)}]{Mehta2005PRA}%
  \BibitemOpen
  \bibfield  {author} {\bibinfo {author} {\bibfnamefont {N.~P.}\ \bibnamefont
  {Mehta}}\ and\ \bibinfo {author} {\bibfnamefont {J.~R.}\ \bibnamefont
  {Shepard}},\ }\href@noop {} {\bibfield  {journal} {\bibinfo  {journal} {Phys.
  Rev. A}\ }\textbf {\bibinfo {volume} {72}},\ \bibinfo {pages} {032728}
  (\bibinfo {year} {2005})}\BibitemShut {NoStop}%
\bibitem [{\citenamefont {Mehta}\ \emph {et~al.}(2007)\citenamefont {Mehta},
  \citenamefont {Esry},\ and\ \citenamefont {Greene}}]{Mehta2007PRA}%
  \BibitemOpen
  \bibfield  {author} {\bibinfo {author} {\bibfnamefont {N.~P.}\ \bibnamefont
  {Mehta}}, \bibinfo {author} {\bibfnamefont {B.~D.}\ \bibnamefont {Esry}}, \
  and\ \bibinfo {author} {\bibfnamefont {C.~H.}\ \bibnamefont {Greene}},\
  }\href {\doibase 10.1103/PhysRevA.76.022711} {\bibfield  {journal} {\bibinfo
  {journal} {Phys. Rev. A}\ }\textbf {\bibinfo {volume} {76}},\ \bibinfo
  {pages} {022711} (\bibinfo {year} {2007})}\BibitemShut {NoStop}%
\bibitem [{\citenamefont {Chuluunbaatar}\ \emph {et~al.}(2006)\citenamefont
  {Chuluunbaatar}, \citenamefont {Gusev}, \citenamefont {Kaschiev},
  \citenamefont {Kaschieva}, \citenamefont {Amaya-Tapia}, \citenamefont
  {Larsen},\ and\ \citenamefont {Vinitsky}}]{Chuluunbaatar2006JPB}%
  \BibitemOpen
  \bibfield  {author} {\bibinfo {author} {\bibfnamefont {O.}~\bibnamefont
  {Chuluunbaatar}}, \bibinfo {author} {\bibfnamefont {A.}~\bibnamefont
  {Gusev}}, \bibinfo {author} {\bibfnamefont {M.}~\bibnamefont {Kaschiev}},
  \bibinfo {author} {\bibfnamefont {V.}~\bibnamefont {Kaschieva}}, \bibinfo
  {author} {\bibfnamefont {A.}~\bibnamefont {Amaya-Tapia}}, \bibinfo {author}
  {\bibfnamefont {S.}~\bibnamefont {Larsen}}, \ and\ \bibinfo {author}
  {\bibfnamefont {S.}~\bibnamefont {Vinitsky}},\ }\href@noop {} {\bibfield
  {journal} {\bibinfo  {journal} {J. Phys. B: At. Mol. Opt. Phys.}\ }\textbf
  {\bibinfo {volume} {39}},\ \bibinfo {pages} {243} (\bibinfo {year}
  {2006})}\BibitemShut {NoStop}%
\bibitem [{\citenamefont {Kartavtsev}\ \emph {et~al.}(2009)\citenamefont
  {Kartavtsev}, \citenamefont {Malykh},\ and\ \citenamefont
  {Sofianos}}]{Kartavtsev2009JETP}%
  \BibitemOpen
  \bibfield  {author} {\bibinfo {author} {\bibfnamefont {O.~I.}\ \bibnamefont
  {Kartavtsev}}, \bibinfo {author} {\bibfnamefont {A.~V.}\ \bibnamefont
  {Malykh}}, \ and\ \bibinfo {author} {\bibfnamefont {S.~A.}\ \bibnamefont
  {Sofianos}},\ }\href@noop {} {\bibfield  {journal} {\bibinfo  {journal}
  {ZhETF}\ }\textbf {\bibinfo {volume} {135}},\ \bibinfo {pages} {419}
  (\bibinfo {year} {2009})}\BibitemShut {NoStop}%
\bibitem [{\citenamefont {Orso}\ \emph {et~al.}(2010)\citenamefont {Orso},
  \citenamefont {Burovski},\ and\ \citenamefont {Jolicoeur}}]{Orso2010PRL}%
  \BibitemOpen
  \bibfield  {author} {\bibinfo {author} {\bibfnamefont {G.}~\bibnamefont
  {Orso}}, \bibinfo {author} {\bibfnamefont {E.}~\bibnamefont {Burovski}}, \
  and\ \bibinfo {author} {\bibfnamefont {T.}~\bibnamefont {Jolicoeur}},\ }\href
  {\doibase 10.1103/PhysRevLett.104.065301} {\bibfield  {journal} {\bibinfo
  {journal} {Phys. Rev. Lett.}\ }\textbf {\bibinfo {volume} {104}},\ \bibinfo
  {pages} {065301} (\bibinfo {year} {2010})}\BibitemShut {NoStop}%
\bibitem [{\citenamefont {Macek}(1968)}]{Macek1968JPB}%
  \BibitemOpen
  \bibfield  {author} {\bibinfo {author} {\bibfnamefont {J.~H.}\ \bibnamefont
  {Macek}},\ }\href@noop {} {\bibfield  {journal} {\bibinfo  {journal}
  {J.~Phys.~B: At. Mol. Opt. Phys.}\ }\textbf {\bibinfo {volume} {1}},\
  \bibinfo {pages} {831} (\bibinfo {year} {1968})}\BibitemShut {NoStop}%
\bibitem [{\citenamefont {Castin}\ \emph {et~al.}(2010)\citenamefont {Castin},
  \citenamefont {Mora},\ and\ \citenamefont {Pricoupenko}}]{Castin2010PRL}%
  \BibitemOpen
  \bibfield  {author} {\bibinfo {author} {\bibfnamefont {Y.}~\bibnamefont
  {Castin}}, \bibinfo {author} {\bibfnamefont {C.}~\bibnamefont {Mora}}, \ and\
  \bibinfo {author} {\bibfnamefont {L.}~\bibnamefont {Pricoupenko}},\ }\href
  {\doibase 10.1103/PhysRevLett.105.223201} {\bibfield  {journal} {\bibinfo
  {journal} {Phys. Rev. Lett.}\ }\textbf {\bibinfo {volume} {105}},\ \bibinfo
  {pages} {223201} (\bibinfo {year} {2010})}\BibitemShut {NoStop}%
\bibitem [{\citenamefont {Guevara}\ \emph {et~al.}(2012)\citenamefont
  {Guevara}, \citenamefont {Wang},\ and\ \citenamefont
  {Esry}}]{Guevara2012PRL}%
  \BibitemOpen
  \bibfield  {author} {\bibinfo {author} {\bibfnamefont {N.~L.}\ \bibnamefont
  {Guevara}}, \bibinfo {author} {\bibfnamefont {Y.}~\bibnamefont {Wang}}, \
  and\ \bibinfo {author} {\bibfnamefont {B.~D.}\ \bibnamefont {Esry}},\ }\href
  {\doibase 10.1103/PhysRevLett.108.213202} {\bibfield  {journal} {\bibinfo
  {journal} {Phys. Rev. Lett.}\ }\textbf {\bibinfo {volume} {108}},\ \bibinfo
  {pages} {213202} (\bibinfo {year} {2012})}\BibitemShut {NoStop}%
\bibitem [{\citenamefont {Petrov}\ \emph {et~al.}(2007)\citenamefont {Petrov},
  \citenamefont {Astrakharchik}, \citenamefont {Papoular}, \citenamefont
  {Salomon},\ and\ \citenamefont {Shlyapnikov}}]{Petrov2007PRL}%
  \BibitemOpen
  \bibfield  {author} {\bibinfo {author} {\bibfnamefont {D.~S.}\ \bibnamefont
  {Petrov}}, \bibinfo {author} {\bibfnamefont {G.~E.}\ \bibnamefont
  {Astrakharchik}}, \bibinfo {author} {\bibfnamefont {D.~J.}\ \bibnamefont
  {Papoular}}, \bibinfo {author} {\bibfnamefont {C.}~\bibnamefont {Salomon}}, \
  and\ \bibinfo {author} {\bibfnamefont {G.~V.}\ \bibnamefont {Shlyapnikov}},\
  }\href {\doibase 10.1103/PhysRevLett.99.130407} {\bibfield  {journal}
  {\bibinfo  {journal} {Phys. Rev. Lett.}\ }\textbf {\bibinfo {volume} {99}},\
  \bibinfo {pages} {130407} (\bibinfo {year} {2007})}\BibitemShut {NoStop}%
\bibitem [{\citenamefont {Olshanii}(1998)}]{Olshanii1998PRL}%
  \BibitemOpen
  \bibfield  {author} {\bibinfo {author} {\bibfnamefont {M.}~\bibnamefont
  {Olshanii}},\ }\href@noop {} {\bibfield  {journal} {\bibinfo  {journal}
  {Phys. Rev. Lett.}\ }\textbf {\bibinfo {volume} {81}},\ \bibinfo {pages}
  {938} (\bibinfo {year} {1998})}\BibitemShut {NoStop}%
\bibitem [{\citenamefont {Bergeman}\ \emph {et~al.}(2003)\citenamefont
  {Bergeman}, \citenamefont {Moore},\ and\ \citenamefont
  {Olshanii}}]{Bergeman2003PRL}%
  \BibitemOpen
  \bibfield  {author} {\bibinfo {author} {\bibfnamefont {T.}~\bibnamefont
  {Bergeman}}, \bibinfo {author} {\bibfnamefont {M.~G.}\ \bibnamefont {Moore}},
  \ and\ \bibinfo {author} {\bibfnamefont {M.}~\bibnamefont {Olshanii}},\
  }\href@noop {} {\bibfield  {journal} {\bibinfo  {journal} {Phys. Rev. Lett.}\
  }\textbf {\bibinfo {volume} {91}},\ \bibinfo {pages} {163201} (\bibinfo
  {year} {2003})}\BibitemShut {NoStop}%
\bibitem [{\citenamefont {Mora}\ \emph
  {et~al.}(2005{\natexlab{a}})\citenamefont {Mora}, \citenamefont {Egger},\
  and\ \citenamefont {Gogolin}}]{Mora2005PRA}%
  \BibitemOpen
  \bibfield  {author} {\bibinfo {author} {\bibfnamefont {C.}~\bibnamefont
  {Mora}}, \bibinfo {author} {\bibfnamefont {R.}~\bibnamefont {Egger}}, \ and\
  \bibinfo {author} {\bibfnamefont {A.~O.}\ \bibnamefont {Gogolin}},\
  }\href@noop {} {\bibfield  {journal} {\bibinfo  {journal} {Phys. Rev. A}\
  }\textbf {\bibinfo {volume} {71}},\ \bibinfo {pages} {052705} (\bibinfo
  {year} {2005}{\natexlab{a}})}\BibitemShut {NoStop}%
\bibitem [{\citenamefont {Mora}\ \emph
  {et~al.}(2005{\natexlab{b}})\citenamefont {Mora}, \citenamefont {Komnik},
  \citenamefont {Egger},\ and\ \citenamefont {Gogolin}}]{Mora2005PRL}%
  \BibitemOpen
  \bibfield  {author} {\bibinfo {author} {\bibfnamefont {C.}~\bibnamefont
  {Mora}}, \bibinfo {author} {\bibfnamefont {A.}~\bibnamefont {Komnik}},
  \bibinfo {author} {\bibfnamefont {R.}~\bibnamefont {Egger}}, \ and\ \bibinfo
  {author} {\bibfnamefont {A.~O.}\ \bibnamefont {Gogolin}},\ }\href {\doibase
  10.1103/PhysRevLett.95.080403} {\bibfield  {journal} {\bibinfo  {journal}
  {Phys. Rev. Lett.}\ }\textbf {\bibinfo {volume} {95}},\ \bibinfo {pages}
  {080403} (\bibinfo {year} {2005}{\natexlab{b}})}\BibitemShut {NoStop}%
\bibitem [{\citenamefont {Levinsen}\ \emph {et~al.}(2014)\citenamefont
  {Levinsen}, \citenamefont {Massignan},\ and\ \citenamefont
  {Parish}}]{Levinsen2014Arxiv}%
  \BibitemOpen
  \bibfield  {author} {\bibinfo {author} {\bibfnamefont {J.}~\bibnamefont
  {Levinsen}}, \bibinfo {author} {\bibfnamefont {P.}~\bibnamefont {Massignan}},
  \ and\ \bibinfo {author} {\bibfnamefont {M.}~\bibnamefont {Parish}},\
  }\href@noop {} {\  (\bibinfo {year} {2014})},\ \Eprint
  {http://arxiv.org/abs/1402.1859} {arXiv:1402.1859 [cond-mat.quant-gas]}
  \BibitemShut {NoStop}%
\bibitem [{\citenamefont {Girardeau}(1960)}]{Girardeau1960JMP}%
  \BibitemOpen
  \bibfield  {author} {\bibinfo {author} {\bibfnamefont {M.~D.}\ \bibnamefont
  {Girardeau}},\ }\href@noop {} {\bibfield  {journal} {\bibinfo  {journal}
  {J.~Math Phys.}\ }\textbf {\bibinfo {volume} {1}},\ \bibinfo {pages} {516}
  (\bibinfo {year} {1960})}\BibitemShut {NoStop}%
\bibitem [{\citenamefont {Brattsev}(1965)}]{Brattsev1965Dokl}%
  \BibitemOpen
  \bibfield  {author} {\bibinfo {author} {\bibfnamefont {V.~F.}\ \bibnamefont
  {Brattsev}},\ }\href@noop {} {\bibfield  {journal} {\bibinfo  {journal}
  {Sov.~Phys.~Dokl.}\ }\textbf {\bibinfo {volume} {10}} (\bibinfo {year}
  {1965})}\BibitemShut {NoStop}%
\bibitem [{\citenamefont {Epstein}(1966)}]{Epstein1966JCP}%
  \BibitemOpen
  \bibfield  {author} {\bibinfo {author} {\bibfnamefont {S.~T.}\ \bibnamefont
  {Epstein}},\ }\href@noop {} {\bibfield  {journal} {\bibinfo  {journal}
  {J.~Chem.~Phys.}\ }\textbf {\bibinfo {volume} {44}},\ \bibinfo {pages} {836}
  (\bibinfo {year} {1966})}\BibitemShut {NoStop}%
\bibitem [{\citenamefont {Starace}\ and\ \citenamefont
  {Webster}(1979)}]{Strace1979PRA}%
  \BibitemOpen
  \bibfield  {author} {\bibinfo {author} {\bibfnamefont {A.~F.}\ \bibnamefont
  {Starace}}\ and\ \bibinfo {author} {\bibfnamefont {G.~L.}\ \bibnamefont
  {Webster}},\ }\href {\doibase 10.1103/PhysRevA.19.1629} {\bibfield  {journal}
  {\bibinfo  {journal} {Phys. Rev. A}\ }\textbf {\bibinfo {volume} {19}},\
  \bibinfo {pages} {1629} (\bibinfo {year} {1979})}\BibitemShut {NoStop}%
\bibitem [{\citenamefont {Coelho}\ and\ \citenamefont
  {Hornos}(1991)}]{Coelho1991PRA}%
  \BibitemOpen
  \bibfield  {author} {\bibinfo {author} {\bibfnamefont {H.~T.}\ \bibnamefont
  {Coelho}}\ and\ \bibinfo {author} {\bibfnamefont {J.~E.}\ \bibnamefont
  {Hornos}},\ }\href {\doibase 10.1103/PhysRevA.43.6379} {\bibfield  {journal}
  {\bibinfo  {journal} {Phys. Rev. A}\ }\textbf {\bibinfo {volume} {43}},\
  \bibinfo {pages} {6379} (\bibinfo {year} {1991})}\BibitemShut {NoStop}%
\bibitem [{\citenamefont {Adhikari}\ \emph {et~al.}(1992)\citenamefont
  {Adhikari}, \citenamefont {Brito}, \citenamefont {Coelho},\ and\
  \citenamefont {Das}}]{Adhikari1992NuovoCimento}%
  \BibitemOpen
  \bibfield  {author} {\bibinfo {author} {\bibfnamefont {S.~K.}\ \bibnamefont
  {Adhikari}}, \bibinfo {author} {\bibfnamefont {V.}~\bibnamefont {Brito}},
  \bibinfo {author} {\bibfnamefont {H.}~\bibnamefont {Coelho}}, \ and\ \bibinfo
  {author} {\bibfnamefont {T.}~\bibnamefont {Das}},\ }\href@noop {} {\bibfield
  {journal} {\bibinfo  {journal} {Il Nuovo Cimento B Series 11}\ }\textbf
  {\bibinfo {volume} {107}},\ \bibinfo {pages} {77} (\bibinfo {year}
  {1992})}\BibitemShut {NoStop}%
\bibitem [{\citenamefont {Gaudin}\ and\ \citenamefont
  {Derrida}(1975)}]{GaudinJdePhys1975}%
  \BibitemOpen
  \bibfield  {author} {\bibinfo {author} {\bibfnamefont {M.}~\bibnamefont
  {Gaudin}}\ and\ \bibinfo {author} {\bibfnamefont {B.}~\bibnamefont
  {Derrida}},\ }\href@noop {} {\bibfield  {journal} {\bibinfo  {journal} {J. de
  Phys.}\ }\textbf {\bibinfo {volume} {36}},\ \bibinfo {pages} {1183} (\bibinfo
  {year} {1975})}\BibitemShut {NoStop}%
\bibitem [{\citenamefont {Kartavstev}\ and\ \citenamefont
  {Malykh}(2007)}]{Kartavtsev2007JPhysB}%
  \BibitemOpen
  \bibfield  {author} {\bibinfo {author} {\bibfnamefont {O.}~\bibnamefont
  {Kartavstev}}\ and\ \bibinfo {author} {\bibfnamefont {A.}~\bibnamefont
  {Malykh}},\ }\href@noop {} {\bibfield  {journal} {\bibinfo  {journal} {J.
  Phys. B: At. Mol. Opt. Phys.}\ }\textbf {\bibinfo {volume} {40}},\ \bibinfo
  {pages} {1429} (\bibinfo {year} {2007})}\BibitemShut {NoStop}%
\bibitem [{\citenamefont {Pricoupenko}\ and\ \citenamefont
  {Pedri}(2010)}]{Pricoupenko2010PRA}%
  \BibitemOpen
  \bibfield  {author} {\bibinfo {author} {\bibfnamefont {L.}~\bibnamefont
  {Pricoupenko}}\ and\ \bibinfo {author} {\bibfnamefont {P.}~\bibnamefont
  {Pedri}},\ }\href {\doibase 10.1103/PhysRevA.82.033625} {\bibfield  {journal}
  {\bibinfo  {journal} {Phys. Rev. A}\ }\textbf {\bibinfo {volume} {82}},\
  \bibinfo {pages} {033625} (\bibinfo {year} {2010})}\BibitemShut {NoStop}%
\bibitem [{\citenamefont {Levinsen}\ and\ \citenamefont
  {Parish}(2013)}]{Levinsen2013PRL}%
  \BibitemOpen
  \bibfield  {author} {\bibinfo {author} {\bibfnamefont {J.}~\bibnamefont
  {Levinsen}}\ and\ \bibinfo {author} {\bibfnamefont {M.~M.}\ \bibnamefont
  {Parish}},\ }\href {\doibase 10.1103/PhysRevLett.110.055304} {\bibfield
  {journal} {\bibinfo  {journal} {Phys. Rev. Lett.}\ }\textbf {\bibinfo
  {volume} {110}},\ \bibinfo {pages} {055304} (\bibinfo {year}
  {2013})}\BibitemShut {NoStop}%
\bibitem [{\citenamefont {Blume}(2012)}]{Blume2012PRL}%
  \BibitemOpen
  \bibfield  {author} {\bibinfo {author} {\bibfnamefont {D.}~\bibnamefont
  {Blume}},\ }\href {\doibase 10.1103/PhysRevLett.109.230404} {\bibfield
  {journal} {\bibinfo  {journal} {Phys. Rev. Lett.}\ }\textbf {\bibinfo
  {volume} {109}},\ \bibinfo {pages} {230404} (\bibinfo {year}
  {2012})}\BibitemShut {NoStop}%
\bibitem [{\citenamefont {Gharashi}\ and\ \citenamefont
  {Blume}(2013)}]{Gharashi2013PRL}%
  \BibitemOpen
  \bibfield  {author} {\bibinfo {author} {\bibfnamefont {S.~E.}\ \bibnamefont
  {Gharashi}}\ and\ \bibinfo {author} {\bibfnamefont {D.}~\bibnamefont
  {Blume}},\ }\href {\doibase 10.1103/PhysRevLett.111.045302} {\bibfield
  {journal} {\bibinfo  {journal} {Phys. Rev. Lett.}\ }\textbf {\bibinfo
  {volume} {111}},\ \bibinfo {pages} {045302} (\bibinfo {year}
  {2013})}\BibitemShut {NoStop}%
\bibitem [{\citenamefont {Sowi\ifmmode~\acute{n}\else \'{n}\fi{}ski}\ \emph
  {et~al.}(2013)\citenamefont {Sowi\ifmmode~\acute{n}\else \'{n}\fi{}ski},
  \citenamefont {Grass}, \citenamefont {Dutta},\ and\ \citenamefont
  {Lewenstein}}]{Sowinski2013PRA}%
  \BibitemOpen
  \bibfield  {author} {\bibinfo {author} {\bibfnamefont {T.}~\bibnamefont
  {Sowi\ifmmode~\acute{n}\else \'{n}\fi{}ski}}, \bibinfo {author}
  {\bibfnamefont {T.}~\bibnamefont {Grass}}, \bibinfo {author} {\bibfnamefont
  {O.}~\bibnamefont {Dutta}}, \ and\ \bibinfo {author} {\bibfnamefont
  {M.}~\bibnamefont {Lewenstein}},\ }\href {\doibase
  10.1103/PhysRevA.88.033607} {\bibfield  {journal} {\bibinfo  {journal} {Phys.
  Rev. A}\ }\textbf {\bibinfo {volume} {88}},\ \bibinfo {pages} {033607}
  (\bibinfo {year} {2013})}\BibitemShut {NoStop}%
\bibitem [{\citenamefont {Lindgren}\ \emph {et~al.}(2013)\citenamefont
  {Lindgren}, \citenamefont {Rotureau}, \citenamefont {Forssén}, \citenamefont
  {Volosniev},\ and\ \citenamefont {Zinner}}]{Lindgren2013Arxiv}%
  \BibitemOpen
  \bibfield  {author} {\bibinfo {author} {\bibfnamefont {E.~J.}\ \bibnamefont
  {Lindgren}}, \bibinfo {author} {\bibfnamefont {J.}~\bibnamefont {Rotureau}},
  \bibinfo {author} {\bibfnamefont {C.}~\bibnamefont {Forssén}}, \bibinfo
  {author} {\bibfnamefont {A.~G.}\ \bibnamefont {Volosniev}}, \ and\ \bibinfo
  {author} {\bibfnamefont {N.~T.}\ \bibnamefont {Zinner}},\ }\href@noop {} {\
  (\bibinfo {year} {2013})},\ \Eprint {http://arxiv.org/abs/1304.2992v1}
  {arXiv:1304.2992v1 [cond-mat]} \BibitemShut {NoStop}%
\bibitem [{\citenamefont {Daily}\ and\ \citenamefont
  {Greene}(2014)}]{Daily2014PRA}%
  \BibitemOpen
  \bibfield  {author} {\bibinfo {author} {\bibfnamefont {K.~M.}\ \bibnamefont
  {Daily}}\ and\ \bibinfo {author} {\bibfnamefont {C.~H.}\ \bibnamefont
  {Greene}},\ }\href {\doibase 10.1103/PhysRevA.89.012503} {\bibfield
  {journal} {\bibinfo  {journal} {Phys. Rev. A}\ }\textbf {\bibinfo {volume}
  {89}},\ \bibinfo {pages} {012503} (\bibinfo {year} {2014})}\BibitemShut
  {NoStop}%
\bibitem [{\citenamefont {Patil}(2009)}]{Patil2009EJP}%
  \BibitemOpen
  \bibfield  {author} {\bibinfo {author} {\bibfnamefont {S.}~\bibnamefont
  {Patil}},\ }\href@noop {} {\bibfield  {journal} {\bibinfo  {journal} {Eur. J.
  Phys.}\ }\textbf {\bibinfo {volume} {30}},\ \bibinfo {pages} {629} (\bibinfo
  {year} {2009})}\BibitemShut {NoStop}%
\end{thebibliography}%

\end{document}